\documentclass[twocolumn]{aastex63}

\usepackage[T1]{fontenc}                 
\usepackage{amsmath,mathtools}           
\usepackage{booktabs,makecell,multirow}  
\usepackage{hyperref}                    

\hypersetup{colorlinks=true,linkcolor=black,citecolor=black,filecolor=black,urlcolor=black}

\newcommand{\tableindent}{\hspace{1em}}

\DeclarePairedDelimiter{\paren}{\lparen}{\rparen}
\DeclarePairedDelimiter{\ave}{\langle}{\rangle}
\DeclarePairedDelimiter{\abs}{\lvert}{\rvert}
\newcommand{\dd}{\mathrm{d}}

\newcommand{\muas}{\mathrm{\mu as}}
\newcommand{\hours}{\mathrm{hr}}
\newcommand{\seconds}{\mathrm{s}}
\newcommand{\ghz}{\mathrm{GHz}}
\newcommand{\microns}{\mathrm{\mu m}}
\newcommand{\kpc}{\mathrm{kpc}}
\newcommand{\jy}{\mathrm{Jy}}
\newcommand{\mjy}{\mathrm{mJy}}
\newcommand{\erg}{\mathrm{erg}}
\newcommand{\gcc}{\mathrm{g\ cm^{-3}}}
\newcommand{\mhz}{\mathrm{mHz}}

\newcommand{\msun}{M_\odot}
\newcommand{\mprot}{m_\mathrm{p}}
\newcommand{\sigmat}{\sigma_\mathrm{T}}

\newcommand{\tti}{T_\mathrm{i}}
\newcommand{\tte}{T_\mathrm{e}}
\newcommand{\rlow}{R_\mathrm{low}}
\newcommand{\rhigh}{R_\mathrm{high}}
\newcommand{\mdotedd}{\dot{M}_\mathrm{Edd}}
\newcommand{\xbar}{\bar{x}}
\newcommand{\ybar}{\bar{y}}
\newcommand{\xbarave}{\bar{x}_\mathrm{ave}}
\newcommand{\ybarave}{\bar{y}_\mathrm{ave}}
\newcommand{\flog}{F_\mathrm{log}}

\newcommand{\code}[1]{\texttt{#1}}

\shorttitle{Effects of Tilt on Time Variability}
\shortauthors{White and Quataert}

\begin{document}

\title{The Effects of Tilt on the Time Variability of Millimeter and Infrared Emission from Sagittarius~A*}
\author{Christopher~J.~White}
\affiliation{Department of Astrophysical Sciences, Princeton University, Peyton Hall, Princeton, NJ 08544, USA}
\author[0000-0001-9185-5044]{Eliot~Quataert}
\affiliation{Department of Astrophysical Sciences, Princeton University, Peyton Hall, Princeton, NJ 08544, USA}

\begin{abstract}

  Using a combination of general-relativistic magnetohydrodynamics simulations and ray tracing of synchrotron emission, we study the effect of modest ($24^\circ$) misalignment between the black hole spin and plasma angular momentum, focusing on the variability of total flux, image centroids, and image sizes. We consider both millimeter and infrared (IR) observables motivated by Sagittarius~A* (Sgr~A*), though our results apply more generally to optically thin flows. For most quantities, tilted accretion is more variable, primarily due to a significantly hotter and denser ``coronal'' region well off the disk midplane. We find (1) a $150\%$ increase in millimeter light curve variability when adding tilt to the flow;\ (2) the tilted image centroid in the millimeter shifts on a scale of $3.7\ \muas$ over $28$ hours ($5000$ gravitational times) for some electron temperature models;\ (3) tilted disk image diameters in the millimeter can be $10\%$ larger ($52$ versus $47\ \muas$) than those of aligned disks at certain viewing angles;\ (4) the tilted models produce significant IR flux, similar to that seen in Sgr~A*, with comparable or even greater variability than observed;\ (5) for some electron models, the tilted IR centroid moves by more than $50\ \muas$ over several hours, in a similar fashion to the centroid motion detected by the GRAVITY interferometer.

\strut

\end{abstract}

\section{Introduction}
\label{sec:introduction}

The environment around and behavior of the supermassive black hole Sagittarius~A* (Sgr~A*) in the galactic center have been the focus of a number of observational efforts in recent years. For example, we now have a census of nearby massive stars, including their motions and mass loss rates \citep{Paumard2006,Lu2009,YusefZadeh2015}; a detection of the relativistic precession of one such star \citep{Gravity2020b}; a measurement of the X-ray spectrum of diffuse plasma around the black hole \citep{Wang2013}; a characterization of the millimeter rotation measure \citep{Bower2018}; a measurement of gas density at a couple thousand gravitational radii \citep{Gillessen2019}; and a detection of horizon-scale motion in the infrared (IR) during a flare \citep{Gravity2018}. With data probing ever smaller scales, a unified picture is beginning to emerge of how material is fed into the system, to what extent it falls into the black hole or is expelled, and which mechanisms generate electromagnetic radiation in each part of the spectrum. The physics of the accretion flow at the scale of the event-horizon, where general relativity (GR) is of the utmost importance, moderates these processes.

With accretion far below the Eddington rate \citep{Marrone2007}, we expect the galactic center to reside in the low-luminosity, geometrically thick regime as described for example in \citet{Narayan1998}, \citet{Blandford1999}, and \citet{Yuan2014}. These radiatively inefficient accretion flows have been the focus of a large body of three-dimensional general-relativistic magnetohydrodynamics simulations --- studying, for example, jet launching in the magnetically arrested disk \citep[MAD,][]{Narayan2003} case \citep{Punsly2009,Tchekhovskoy2011,McKinney2012}, pair production \citep{Moscibrodzka2011}, quasi-periodic oscillations \citep{Dolence2012}, long-term steady states \citep{Narayan2012,White2020a}, outflows \citep{Sadowski2013,Yuan2015}, electron thermodynamics \citep{Ressler2015,Sadowski2017,Chael2017,Ryan2018,Dexter2020a}, pressure anisotropies and thermal conductivities \citep{Foucart2017}, and dynamos contributing to jets \citep{Liska2020} --- using a variety of codes as compared in \citet{Porth2019}.

However, the distribution of plasma around Sgr~A* and other similar systems might be more complicated than many of our models. The vast majority of simulations are initialized with a largely rotationally supported torus in hydrostatic equilibrium, whose angular momentum is parallel to the black hole spin, if the black hole is even taken to be spinning at all. There is mounting numerical evidence, however, that the feeding of this particular black hole via stellar winds results in a more disordered flow at small radii, with angular momentum not constrained to follow black hole spin \citep{Ressler2020a,Ressler2020b}.

This misalignment is the essence of tilted disks, where the strongly radially dependent Lense--Thirring torque differentially precesses the material, leading to a complex shape. Early simulations of tilted disks confirmed their nonplanar nature \citep{Fragile2005,Fragile2007}, and indicated they could contain a pair of standing shocks \citep{Fragile2008,Generozov2014}. With sufficient misalignment and a large black hole spin, the effect of these shocks results in order-unity changes to the heating and entropy generation in the accretion flow \citep{White2019}. Tilted disks can also have a large effect on the direction of the jet \citep{Liska2018}. A particular concern is that a tilted disk in nature modeled with strictly aligned simulations could easily bias interpretations of horizon-scale observations, such as the inferred black hole mass or constraints on alternate theories of gravity \citep{White2020b}. We therefore seek to better understand these effects, especially in the context of GRAVITY and Event Horizon Telescope (EHT) observations.

In the near future, the EHT will produce the first set of resolved horizon-scale images of Sgr~A*. This follows on the successful observations of the black hole at the center of M87 \citep{EHT2019a}. The two sources have approximately the same horizon size on the sky, and they are both low-luminosity systems. Thus their general observational characteristics should be similar, and in particular one might expect measurements of Sgr~A* to be of a quality similar to those of M87.

However, the picture is complicated by a difference in timescales for the two systems. The mass of the Sgr~A* black hole has been confirmed by numerous measurements to be a little above $4 \times 10^6\ \msun$ \citep{Ghez2003,Ghez2008,Gillessen2009,Gravity2019}, whereas estimates of M87's mass range from $3 \times 10^9\ \msun$ \citep{Walsh2013} to $6 \times 10^9\ \msun$ \citep{Gebhardt2011,EHT2019f}. Dynamical times are thus several orders of magnitude shorter in the former, where indeed $G M / c^3 = 20\ \seconds$. Interpreting sparse interferometric data, taken with baselines changing over the course of several hours, of a source that is itself potentially rapidly varying, presents a unique challenge.

The combination of time variability and possible nonaligned accretion might seem discouraging for interpreting existing and forthcoming data from EHT, GRAVITY, and other instruments. However, it is possible that the one effect can be used to probe the other. Here we investigate the difference between aligned and tilted disks in the context of how they induce time-variable signatures in observable quantities. This work is a natural complement to our earlier consideration of mostly time-averaged properties of aligned versus tilted resolved images \citep{White2020b}. In this case we limit our attention to Sgr~A*, where such variability should manifest on timescales of minutes to hours.

Other studies have found that tilted disks in general tend to display more variability in their images. \Citet{Dexter2013} showed that a disk tilted by $15^\circ$ can lead to variability at both millimeter and IR wavelengths consistent with observations of Sgr~A*. Their simulations evolved an internal rather than total energy equation, and thus did not capture magnetic dissipation at the grid scale the way we can now. More recently, \citet{Chatterjee2020} showed how a number of properties changed with tilt angle from $0^\circ$ to $30^\circ$ to $60^\circ$. Their variability results focused on M87, with tilt causing larger fluctuations in the millimeter light curve.

Our analysis takes into account over $96{,}000$ images generated via ray tracing, accounting for optically thin synchrotron emission appropriate to Sgr~A*. We consider millimeter images at $230\ \ghz$, relevant for the EHT; IR images at $2.2\ \microns$, corresponding to GRAVITY observations; and several additional frequencies used to capture spectral energy distributions.

In what follows, we show that even simple measurements are sensitive to tilt. Misaligned angular momentum can reveal itself in the time variability of a number of observables. Here we focus on elucidating these trends and understanding their physical origin. Applying lessons learned to raw interferometric data, or even passing ideal models through telescope pipelines, is beyond the scope of this work. Thus we present results that, while commensurate with the abilities of EHT and GRAVITY, are framed in a partially instrument-agnostic way. More EHT-specific studies of variability in observed closure phases and visibility amplitudes have already been respectively performed by, e.g., \citet{Medeiros2017} and \citet{Medeiros2018}, though not yet in the context of tilted disks.

One would ideally have a large corpus of simulations (especially those run for very long times, as in \citet{Chan2015a}, for example) with different tilt, magnetic flux, and black hole spin in order to densely cover the relevant parameter space and to guard against being mislead by sample variance. Our initial results here are based only on two simulations, but their purpose is not to decide whether any given accretion flow is tilted with absolute certainty, but rather to provide a set of expectations for when and how tilt may affect observables. Furthermore, we endeavor to provide physical reasoning to explain our numerical observations, connecting the effects we see to well-established qualitative differences between tilted and aligned disks.

We describe the numerical details of our simulations and ray tracing in Section~\ref{sec:numerics}. We proceed to describe how tilt affects low-order moments of $230\ \ghz$ images (Section~\ref{sec:230ghz_moments}), smaller-scale features in those images (Section~\ref{sec:230ghz_features}), and the moments of $2.2\ \microns$ observations (Section~\ref{sec:2micron_moments}). Section~\ref{sec:discussion} synthesizes and summarizes these results.

\section{Numerical Procedure}
\label{sec:numerics}

We base our investigation on the two GRMHD simulations employed in \citet{White2020b}. In summary, they are both initialized with a torus around a black hole with spin $a = 0.9$, in one case with aligned angular momentum and in the other case with a misalignment of $24^\circ$. The tori are based on the prescription of \citet{Fishbone1976}, with adiabatic index $\Gamma = 4/3$, inner edge at $r = 15\ G M / c^2$, and pressure maximum at $r = 25\ G M / c^2$. The initial magnetic field consists of a single set of concentric loops in the poloidal plane of the torus, with a density-weighted average plasma $\beta^{-1}$ of $0.01$.

The simulations employ a spherical $r \times \theta \times \phi$ static mesh refinement grid with an effective resolution of $448 \times 256 \times 352$ within $50^\circ$ of the midplane. The radial extent is from $0.926$ times the horizon radius to $r = 100\ G M / c^2$, with cells spaced logarithmically ($239$ cells per decade in radius). The simulations are evolved to a time of $t = 11{,}000\ G M / c^3$ with the GR capabilities \citep{White2016} of the \code{Athena++} code \citep{Stone2020}. By this point magnetorotational turbulence has fully developed, the inner regions important for synchrotron emission have reached inflow equilibrium, and the tilted disk has acquired a characteristic warped and twisted shape. This steady state is established out beyond $r = 10\ G M / c^2$ (for the tilted disk) or $r = 20\ G M / c^2$ (for the aligned disk), corresponding to angular separations of $50\text{--}100\ \muas$ for Sgr~A*. The emitting region is comparable to or smaller in size than this. The dimensionless magnetic flux
\begin{equation}
  \varphi = \frac{c^{3/2}}{2 G M \dot{M}^{1/2}} \oint \abs{B^r} \sqrt{-g} \, \dd\theta \, \dd\phi
\end{equation}
(in Gaussian units) saturates between $7$ and $17$, making these standard and normal evolution (SANE) accretion flows, as opposed to MAD flows with $\varphi \sim 47$ \citep{Tchekhovskoy2011}. Our analysis uses only data from $t \geq 6000\ G M / c^3$ (henceforth considered the zero point in time), with snapshots taken every $10\ G M / c^3$.

Each snapshot is used as input to the GR ray tracing code \code{grtrans} \citep{Dexter2009,Dexter2016}, as was also done in \citet{White2020b}. We choose parameters appropriate for Sgr~A*:\ a mass of $M = 4.152 \times 10^6\ \msun$ and a distance of $8.178\ \kpc$ \citep{Gravity2019}.

For this study we focus on images as they would appear from a line of sight near to the (north) pole of the black hole, specifically placing the camera at a polar angle (colatitude) $\theta$ of $5^\circ$, $10^\circ$, $20^\circ$, or $45^\circ$. This is motivated by GRAVITY observations of nearly horizon-scale motion of the image centroid that are suggestive of a hot spot circling Sgr~A* in a plane close to orthogonal to our line of sight \citep{Gravity2018}. We note, however, that the viewing angle and interpretation are by no means settled by this one set of observations \citep[see, e.g.,][]{Matsumoto2020}, and indeed inclinations $\theta \leq 20^\circ$ were disfavored by early variability considerations in \citet{Dexter2010}. Should Sgr~A* prove to be essentially edge-on, the analysis here would need to be redone for quantitative application to Sgr~A*.

When imaging the aligned disk, the azimuthal viewing angle $\phi$ should not affect our statistics, and so we set it to the same $0^\circ$ in all cases, where the initial disk angular momentum is pointed toward $\theta = 24^\circ$, $\phi = 0^\circ$. For the tilted disks, we always consider the four azimuths $0^\circ$, $90^\circ$, $180^\circ$, and $270^\circ$. In all cases the camera is placed at a radius of $100\ G M / c^2$.

The underlying simulations are ideal, single-fluid GRMHD calculations, and so we must have a prescription to convert gas density and pressure into the electron temperature needed to calculate synchrotron emissivities. We adopt the parameterized prescription of \citet{Moscibrodzka2016}, where the ion-to-electron temperature ratio is a function of plasma $\beta$:
\begin{equation}
  \frac{\tti}{\tte} = \frac{\rlow + \beta^2 \rhigh}{1 + \beta^2}.
\end{equation}
Given the uncertainties in how hot the electrons in such plasmas are (and even whether this sort of post-processing is sufficient to capture their temperatures), we adopt several instances of this prescription covering a range of plausible values. In some cases we set $\rlow = 1$ and either $\rhigh = 40$ or $\rhigh = 80$. For reference, the image library initially used by EHT to model M87 consists entirely of electron models in this form, with $\rlow = 1$ and $1 \leq \rhigh \leq 160$. These models are motivated by turbulent heating in which electrons are preferentially heated at low magnetizations, and they are similar in this regard to the more sophisticated in situ \citet[``H10'']{Howes2010} and \citet[``K19'']{Kawazura2019} models explored in \citet{Dexter2020a}. In that work, aligned SANE simulations similar to ours were found to broadly reproduce the observed millimeter to IR spectrum of Sgr~A* when employing such models. We also consider two constant temperature models:\ $\tti / \tte, \rlow, \rhigh = 4$ and $\tti / \tte, \rlow, \rhigh = 8$. These are similar in spirit to the reconnection-inspired \citet[``R17'']{Rowan2017} and \citet[``W18'']{Werner2018} electron evolution models used in \citeauthor{Dexter2020a}, which reproduced the spectra and Faraday rotation of Sgr~A* when applied to MAD simulations.

When ray tracing at $230\ \ghz$ and, for the aligned disk, at $2.2\ \microns$, we use $256^2$ rays covering an image $24\ G M / c^2$ on each side. As we will discuss, the tilted disk tends to have significant IR emission at larger radii, and so for those cases we use $512^2$ rays and an image size of $(48\ G M / c^2)^2$. For comparison, this resolution of $10.7$ pixels per $G M / c^2$ lies between the two highest resolutions ($8$ and $32$ pixels per $G M / c^2$) used in the variability study of \citet{Chan2015b}.\footnote{The statistics we calculate change negligibly if we instead use half the linear resolution.}

We apply a cut in plasma $\sigma$ such that material with $\sigma > 1$ is treated as vacuum for the purposes of ray tracing. This is a standard procedure meant to account for numerical floors artificially mass-loading jets. It is not clear how warranted such a cut is for tilted disks, as we discuss further in Section~\ref{sec:2micron_moments:light_curves}.

With the above settings chosen, there is only one free parameter, which is effectively the physical density scale in the scale-free simulation. We adjust this for each simulation, viewing angle, and electron model until the time-averaged $230\ \ghz$ flux is within $1\%$ of a fiducial value of $2.4\ \jy$.\footnote{This is the value reported in \citet{Doeleman2008}. Though the source certainly varies by more than $1\%$, we enforce agreement at this level in order to make fair comparisons between sets of images.} For the aligned disk, this corresponds to the initial peak density in the torus lying between $5.7 \times 10^{-17}\ \gcc$ and $6.6 \times 10^{-16}\ \gcc$, depending on viewing angle and electron model. These values in turn correspond to a time-averaged accretion rate ranging from $2.9 \times 10^{-8}\ \mdotedd$ to $3.3 \times 10^{-7}\ \mdotedd$, where we define $\mdotedd = 10 \cdot 4 \pi G M \mprot / c \sigmat$. For the tilted disk, the initial peak density ranges from $4.4 \times 10^{-17}\ \gcc$ to $1.5 \times 10^{-16}\ \gcc$, and the accretion rate ranges from $4.9 \times 10^{-8}\ \mdotedd$ to $1.7 \times 10^{-7}\ \mdotedd$. These accretion rates are all reasonable for the low-luminosity system we are studying.

\begin{figure*}
  \centering
  \includegraphics{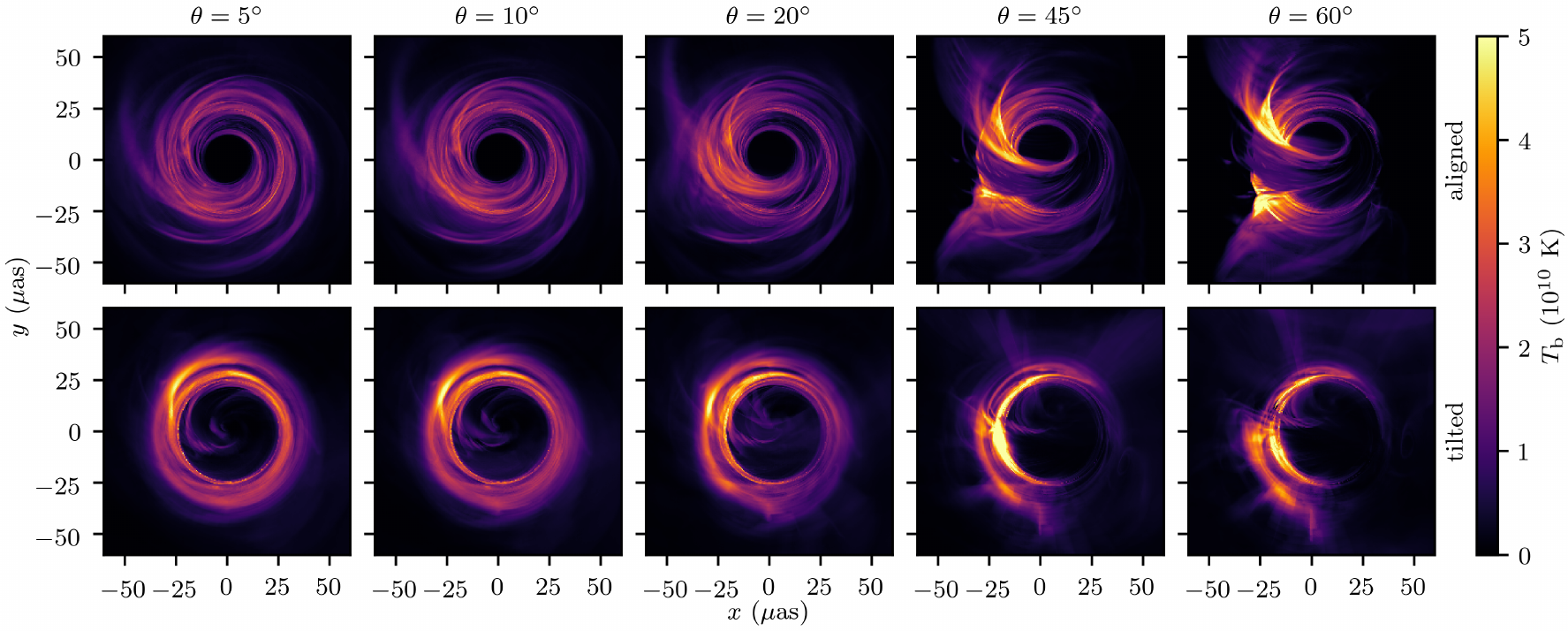}
  \caption{$230\ \ghz$ images made at $t = 3000\ G M / c^3$ with $\rlow = 1$, $\rhigh = 40$, and $\phi = 0^\circ$. There are qualitative differences between the low inclinations ($\theta \leq 20^\circ$) that are the focus of this study and higher inclinations. We include the $45^\circ$ case in our full analysis, and report some statistics for the very similar $60^\circ$ case. \label{fig:images_inclinations}}
\end{figure*}

Illustrative $230\ \ghz$ images are shown in Figure~\ref{fig:images_inclinations}, where the electron model is fixed to $\rlow = 1$, $\rhigh = 40$ and the azimuth is set to $0^\circ$. The snapshots are all taken at the same simulation time, and they show a clear dichotomy between well-defined face-on rings for $\theta \leq 20^\circ$ and less symmetric crescent structures for greater inclinations. The extra $60^\circ$ case is included to show the extent to which a range of mid-latitude camera inclinations produce similar images.

Finally, we define our statistical procedure. A number of physically meaningful quantities (e.g.\ total flux or measured ring size) will have a distribution sampled by each ray-traced image. Ignoring the time-ordering of these samples, we often will summarize the distribution in terms of its mean or standard deviation as inferred from the sample values. In all such cases we follow the maximum-entropy approach of \citet{Oliphant2006}, which places a uniform prior on the mean and a Jeffrey's prior on the standard deviation, to calculate the mean of the posterior for the given statistic. That is, we report the mean of the posterior for the mean, or else the mean of the posterior for the standard deviation.

\section{Image Moments at \texorpdfstring{$230\ \ghz$}{230 GHz}}
\label{sec:230ghz_moments}

We begin by considering $230\ \ghz$ images, but not yet in high resolution. Even the zeroth and first moments of the images (i.e.\ the light curves and centroid locations) convey information about the tilt of the accretion disk.

\subsection{Light Curves}
\label{sec:230ghz_moments:light_curves}

\begin{figure}
  \centering
  \includegraphics{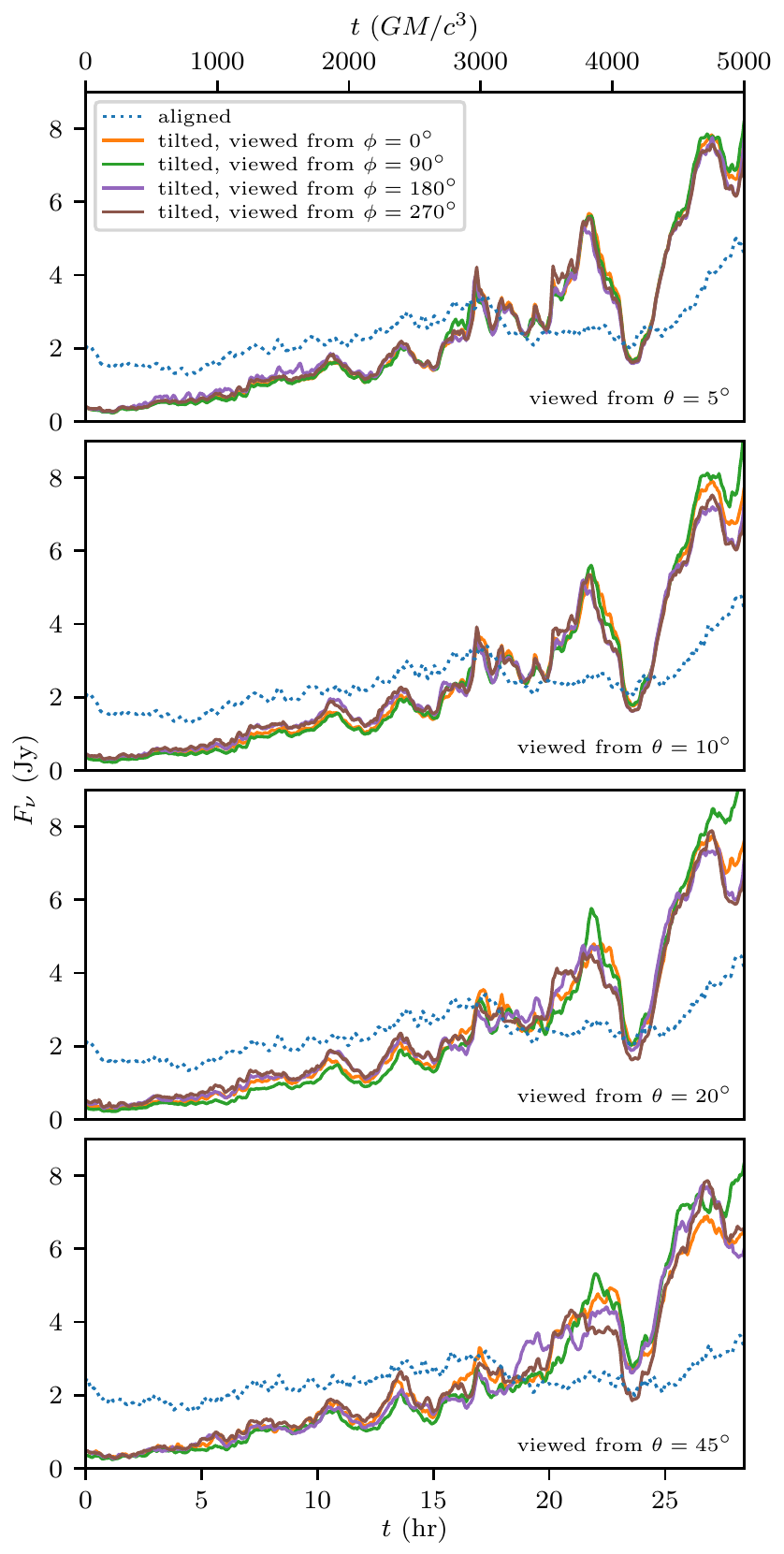}
  \caption{$230\ \ghz$ light curves for the $\rlow = 1$, $\rhigh = 40$ electron model. The tilted simulation shows more variability in total flux than the aligned simulation at all viewing angles studied. \label{fig:light_curves_230ghz}}
\end{figure}

The light curves, defined as
\begin{equation}
  F_\nu = \int\limits_\mathrm{image} I_\nu \, \dd\Omega
\end{equation}
as a function of time, for both simulations at all viewing angles are shown in Figure~\ref{fig:light_curves_230ghz}. There is a secular trend in the tilted curves, but this is the result of long-term variations in the underlying simulation;\ we do not consider this to be physically meaningful time variability. It is more important that all $16$ tilted curves show short-term ($\sim 1\ \hours$) variability with larger amplitudes than in the aligned curves.

Figure~\ref{fig:light_curves_230ghz} also shows a tight correlation between tilted curves of the same viewing inclination but different azimuths. While we know that azimuth can play no role in total light received for $\theta = 0$, this dataset shows that even at $\theta = 45^\circ$ viewing azimuth matters little. This can be reconciled with the lack of axial symmetry in the system by noting that most lines of sight are optically thin, allowing the entire volume to contribute to the image no matter the viewing angle. As a result, changing the camera location shows largely the same integrated emission.

The other three electron models show the same properties:\ larger variations in the tilted case, and a tight correlation between curves that differ only in azimuthal viewing angle.

\begin{figure*}
  \centering
  \includegraphics{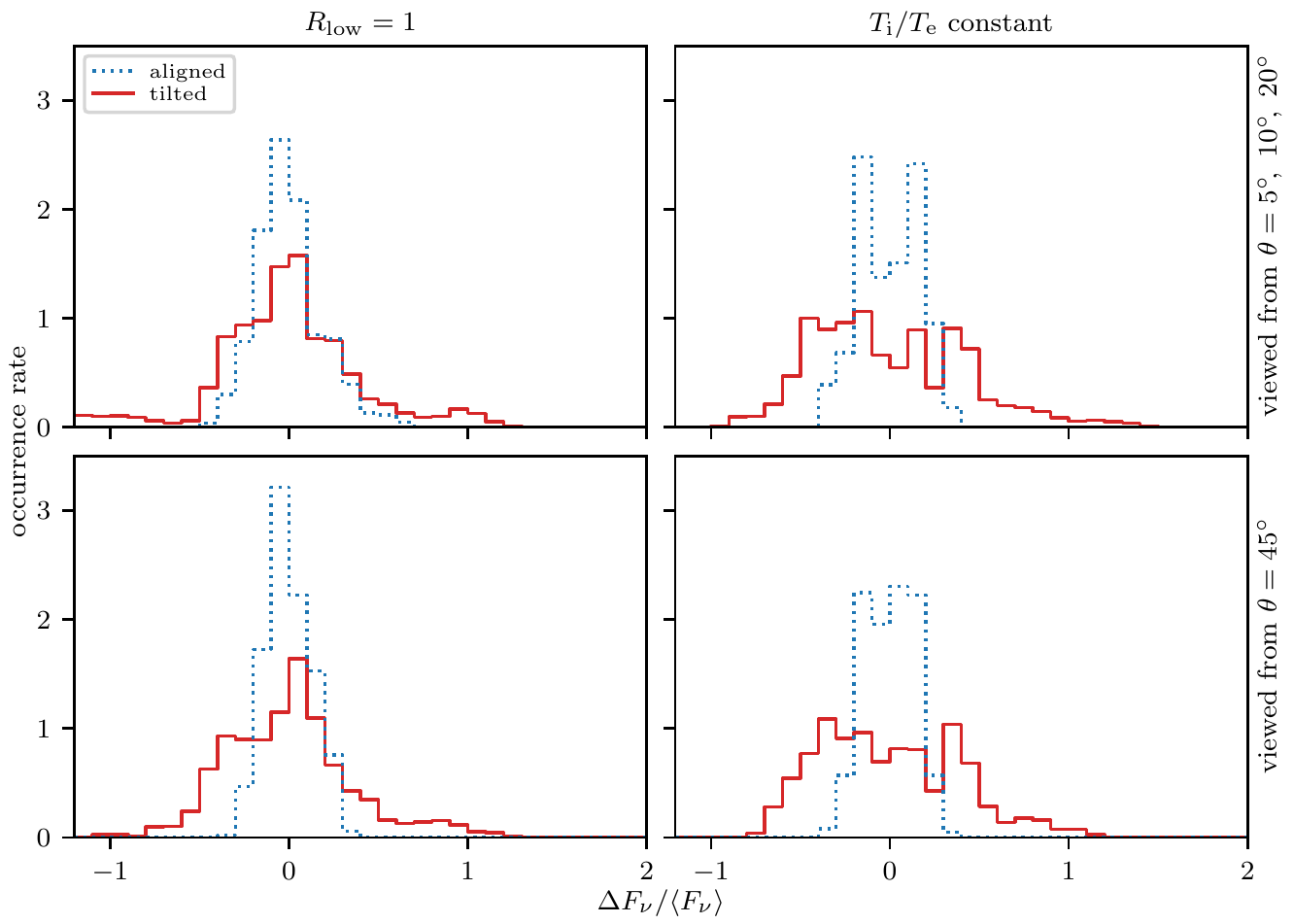}
  \caption{Distributions of $230\ \ghz$ light curve residuals after subtracting the linear trend, partitioned based on the electron model and whether the viewing angle is close to the spin axis. With all electron models and viewing angles studied, tilting a disk results in more variability in the total flux received. \label{fig:light_curve_residuals_230ghz}}
\end{figure*}

In order to quantify variability, we calculate a least-squares linear trend for each light curve individually, subtract this trend to obtain the residuals $\Delta F_\nu$, and normalize the residuals by the time average $\ave{F_\nu}$ (essentially $2.4\ \jy$ in all cases by construction).\footnote{We have experimented with a number of detrending procedures, finding this to capture what we consider meaningful variability while excluding features induced by the finite nature of the initial conditions. Removing Fourier modes with frequencies below approximately $(1700\ G M / c^3)^{-1}$ produces similar results. Decisions regarding detrending should be kept in mind when comparing to others' datasets and analyses.} We then examine the distributions of normalized residuals. As expected from the raw light curves, many of these distributions are similar. In particular, they do not correlate with viewing azimuth, the nearly face-on viewing inclinations ($\theta \leq 20^\circ$) are very similar, and the electron models naturally divide between those with $\rlow = 1$ and those with $\tti / \tte$ constant. Grouping the residuals based on these observations, we plot the corresponding distributions in Figure~\ref{fig:light_curve_residuals_230ghz}.

In all cases, the tilted disks have broader distributions than their aligned counterparts. This is true of all individual light curves, not just the merged datasets. That is, the tilted disk light curves have systematically larger amplitudes of short-term variability.

More quantitatively, when we collect all the normalized residuals for the aligned cases (using all four electron models and all viewing inclinations not exceeding $45^\circ$), the standard deviation is $0.17$. Combining the tilted data, we find $0.40$. The widths of the residual distributions are clearly different, with the tilted central value being $2.4$ times the aligned value. Even at $60^\circ$, the $\rlow = 1$, $\rhigh = 40$ model has a residual width distribution $2.4$ times wider for the tilted disk.

\subsection{Centroid Positions}
\label{sec:230ghz_moments:centroid_positions}

\begin{figure}
  \centering
  \includegraphics{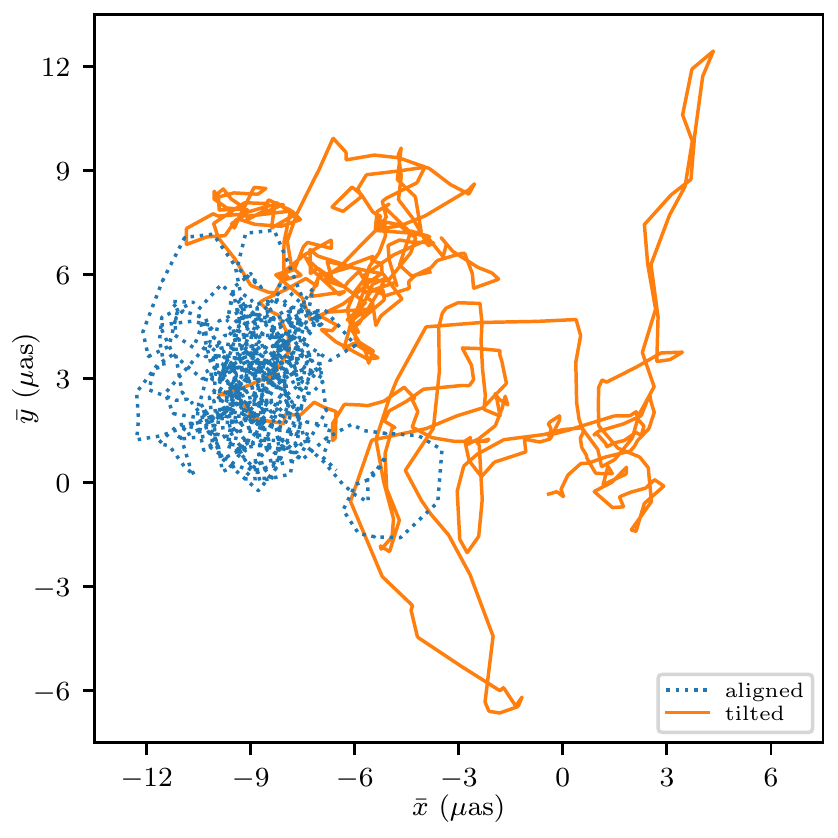}
  \caption{Paths taken by centroids of $230\ \ghz$ light on the sky for the aligned disk and the tilted disk. In both cases we adopt $\rlow = 1$, $\rhigh = 40$, and $\theta = 20^\circ$, though the general trend of greater displacements in the tilted case is true for $\rhigh = 80$ and all inclinations. \label{fig:centroids_230ghz}}
\end{figure}

The trend of greater variability in the tilted disk extends from the zeroth to the first moments of the image, at least in most cases. Define the centroid to be at $(\xbar, \ybar)$, with
\begin{equation}
  \xbar = \frac{1}{F_\nu} \int\limits_\mathrm{image} I_\nu x \, \dd\Omega
\end{equation}
and $\ybar$ defined similarly. The motion of the centroids on the sky for two illustrative cases, one aligned and one tilted, is shown in Figure~\ref{fig:centroids_230ghz}. Here we fix $\rlow = 1$, $\rhigh = 40$, and $\theta = 20^\circ$, and we choose a single $\phi = 0^\circ$ for the tilted case. The other $\rlow = 1$ cases not shown are qualitatively similar:\ the centroid in the tilted case takes larger excursions, while the aligned centroid is more confined. The constant temperature ratio models, however, show less differences with tilt. 

\begin{figure}
  \centering
  \includegraphics{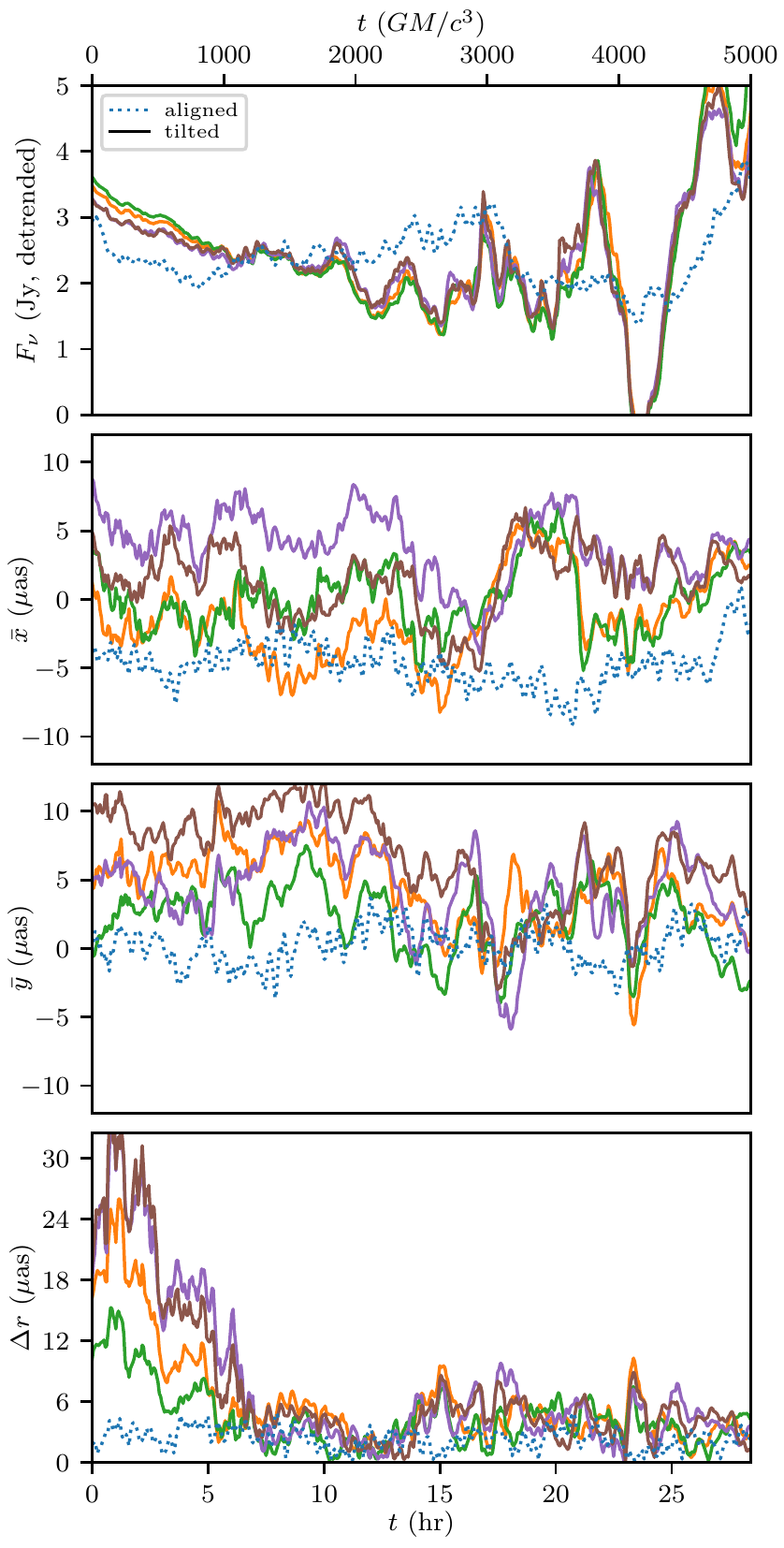}
  \caption{Instantaneous linearly detrended flux, centroid $x$- and $y$-positions, and displacements of instantaneous centroids from centroids of time-averaged images, all at $230\ \ghz$. These curves show the results for $\rlow = 1$, $\rhigh = 40$, and $\theta = 10^\circ$, with other cases appearing similar. Images are rotated to account for varying $\phi$ values in the tilted case, leading to tight correlations in $\xbar$ and $\ybar$. Note $\Delta r$ is systematically larger for the tilted case. \label{fig:centroid_positions_230ghz}}
\end{figure}

We can further quantify this positional variability by examining the distribution of displacements from a central point. Figure~\ref{fig:centroid_positions_230ghz} shows time series of $\xbar$ and $\ybar$, as well as the displacement of the centroid at a given time from the centroid of the time-averaged image. That is, we define
\begin{equation}
  \xbarave = \frac{1}{\ave{F_\nu}} \int\limits_\text{image} \ave{I_\nu} x \, \dd\Omega,
\end{equation}
and similarly for $\ybarave$, where angled brackets denote a time average. Then we define the displacement from the average as
\begin{equation}
  \Delta r = \paren[\big]{(\xbar - \xbarave)^2 + (\ybar - \ybarave)^2}^{1/2}.
\end{equation}
In this case we show the data for $\theta = 10^\circ$.

In the figure, $\xbar$ and $\ybar$ are rotated about the origin to account for the rotation of the camera as $\phi$ varies,\footnote{\code{grtrans} keeps the north pole of the simulation at the top of the image.} and with this adjustment we again see tight correlation between the different viewing azimuths. There is also a clear trend of $\Delta r$ being systematically larger when the disk is tilted. The large excursions of the centroid in these cases are not clearly correlated with flux increases.

\begin{figure*}
  \centering
  \includegraphics{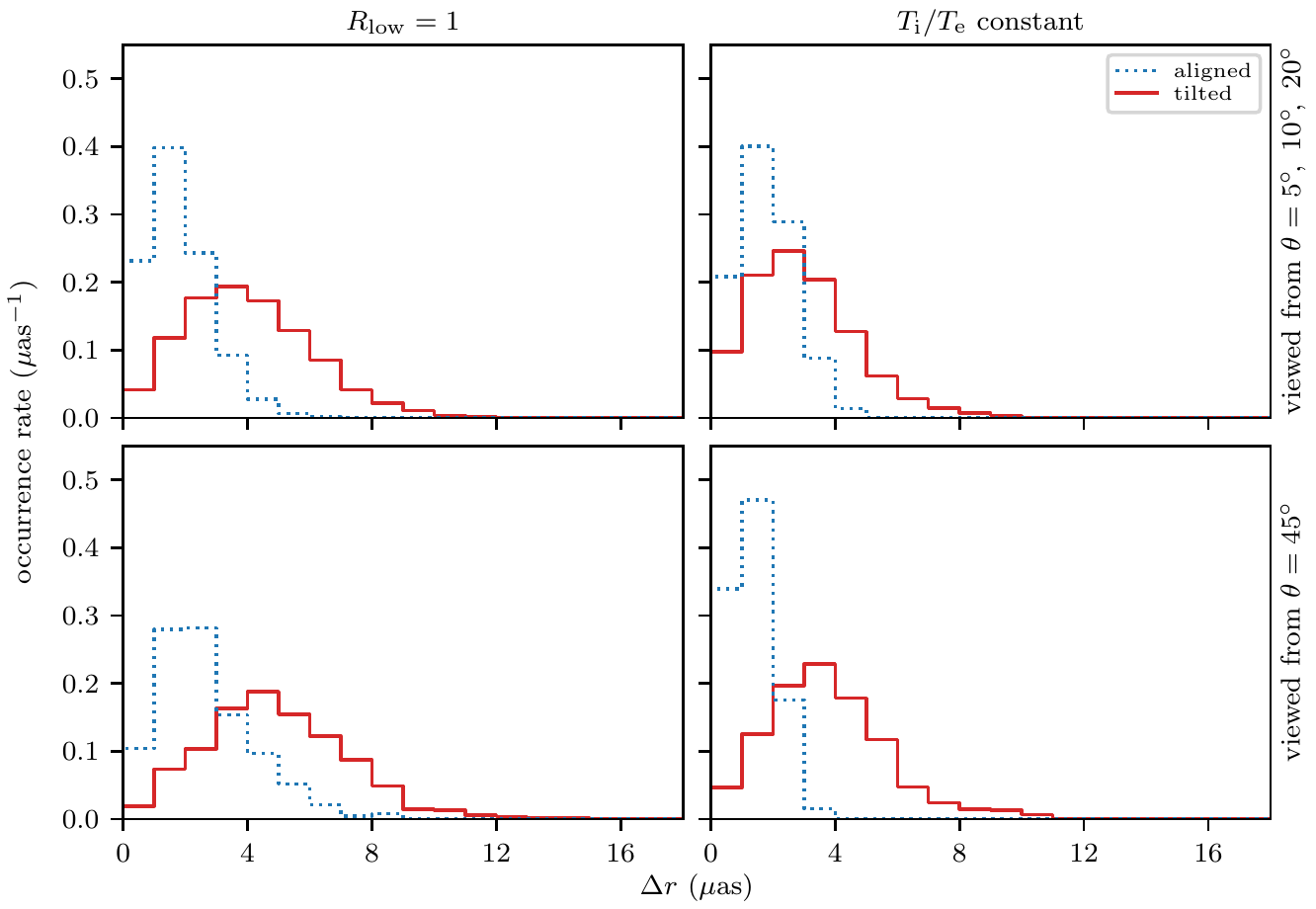}
  \caption{Distributions of displacements of instantaneous centroids from centroids of time-averaged images at $230\ \ghz$. Results are partitioned based on type of electron model and whether the viewing angle is close to parallel with the black hole spin. The tilted disk centroids show systematically larger deviations in all cases. \label{fig:centroid_displacements_230ghz}}
\end{figure*}

The distributions of $\Delta r$ values for all cases are shown in Figure~\ref{fig:centroid_displacements_230ghz}. As with the light curves, we group the data based on electron model ($\rlow = 1$ versus constant temperature ratio) and viewing inclination ($\theta \leq 20^\circ$ versus $\theta = 45^\circ$). In general, the tilted disk instantaneous centroids have larger excursions than those of the aligned disk. Combining the $\Delta r$ values across all our models yields a distribution with mean $1.9\ \muas$ for the aligned cases and $3.7\ \muas$ for the tilted cases;\ tilt results in instantaneous displacements $2.0$ times further from the average image center. We note that at $\theta = 60$, with $\rlow = 1$ and $\rhigh = 40$, the mean $\Delta r$ values are $3.1\ \muas$ (aligned) and $5.4\ \muas$ (tilted), a factor of $1.7$ different.

\section{Image Features at \texorpdfstring{$230\ \ghz$}{230 GHz}}
\label{sec:230ghz_features}

Next we consider spatially resolved $230\ \ghz$ images and their variability.

\subsection{Ring Diameter}
\label{sec:230ghz_features:ring_diameter}

One of the most important measurable properties of a resolved image of horizon-scale accretion is the size of the ring in the image. For example, the EHT Collaboration measured a diameter of $42 \pm 3\ \muas$ for M87 \citep{EHT2019f}. While this is not exactly the diameter on the sky of the photon ring proper, due to emission from other parts of the accretion flow as well as limited angular resolution, the two physical scales are likely related by an order-unity factor.

Though the size of the photon ring proper will not change over the course of observations, the observed ring might still be variable. We investigate this possibility by measuring the ring size frame by frame in our ray-traced simulation images as follows, essentially following the same procedure as in \citet{White2020b}. First we apply a Gaussian blur with a full width at half maximum of $20\ \muas$, in order to approximate the effective resolution of the EHT in the image plane \citep{EHT2019b}. Next we resample the intensity onto a polar grid centered at the origin with $64$ points every $1\ \muas$ in radius and $128$ points in azimuth. The peak brightness is located in each radial spoke, defining a ridgeline circling the origin. We then repeat this process but starting from the geometric center of the interior of the ridgeline rather than the origin, in order to account for any offset the ring might have from the center. The ridgeline is again found, with each radial spoke of the polar grid yielding a measurement of the radius.

We note that while this procedure works well for images of the aligned disk when viewed nearly face-on, often the effects of complex tilted disk geometry or inclined viewing angles are such that some radial spokes do not intersect a well-defined ridgeline. That is, the intensity of the blurred image decreases monotonically moving outward. This happens to approximately $3\%$ of the radial spokes in the tilted images for $\theta \leq 20^\circ$, and to approximately $9\%$ of the spokes in all images at $\theta = 45^\circ$. We simply omit these values when calculating the average ring radius in an image.

\begin{figure}
  \centering
  \includegraphics{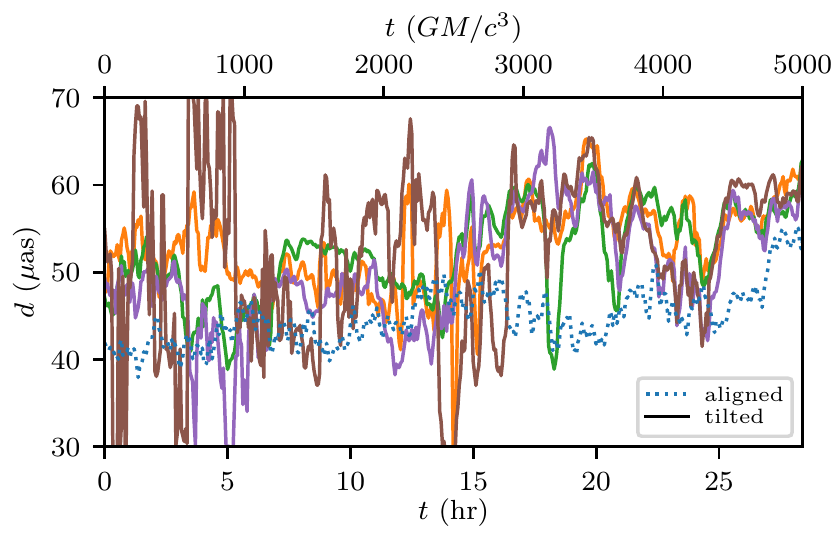}
  \caption{Measured diameters of the blurred ring on the sky at $230\ \ghz$ for several models. These curves show the results for $\rlow = 1$, $\rhigh = 40$, and $\theta = 20^\circ$, with other cases appearing similar. The four tilted curves are for different viewing azimuths $\phi$. The tilted disks display more variability on roughly hour-long timescales than their aligned counterparts. \label{fig:ring_diameters_230ghz}}
\end{figure}

As was already reported in \citet{White2020b}, for nearly face-on images of Sgr~A*, adding tilt can increase the observed size of the ring by $20\%$ and can double a measure of noncircularity;\ other viewing angles and models can be affected even more. Here we focus on the time-variable nature of this measurement. The measured ring diameters as functions of time are shown in Figure~\ref{fig:ring_diameters_230ghz} for $\rlow = 1$, $\rhigh = 40$, and $\theta = 20^\circ$. The other cases are essentially the same, with the aligned ring diameter varying much less than the diameter measured in the tilted case.

\begin{figure*}
  \centering
  \includegraphics{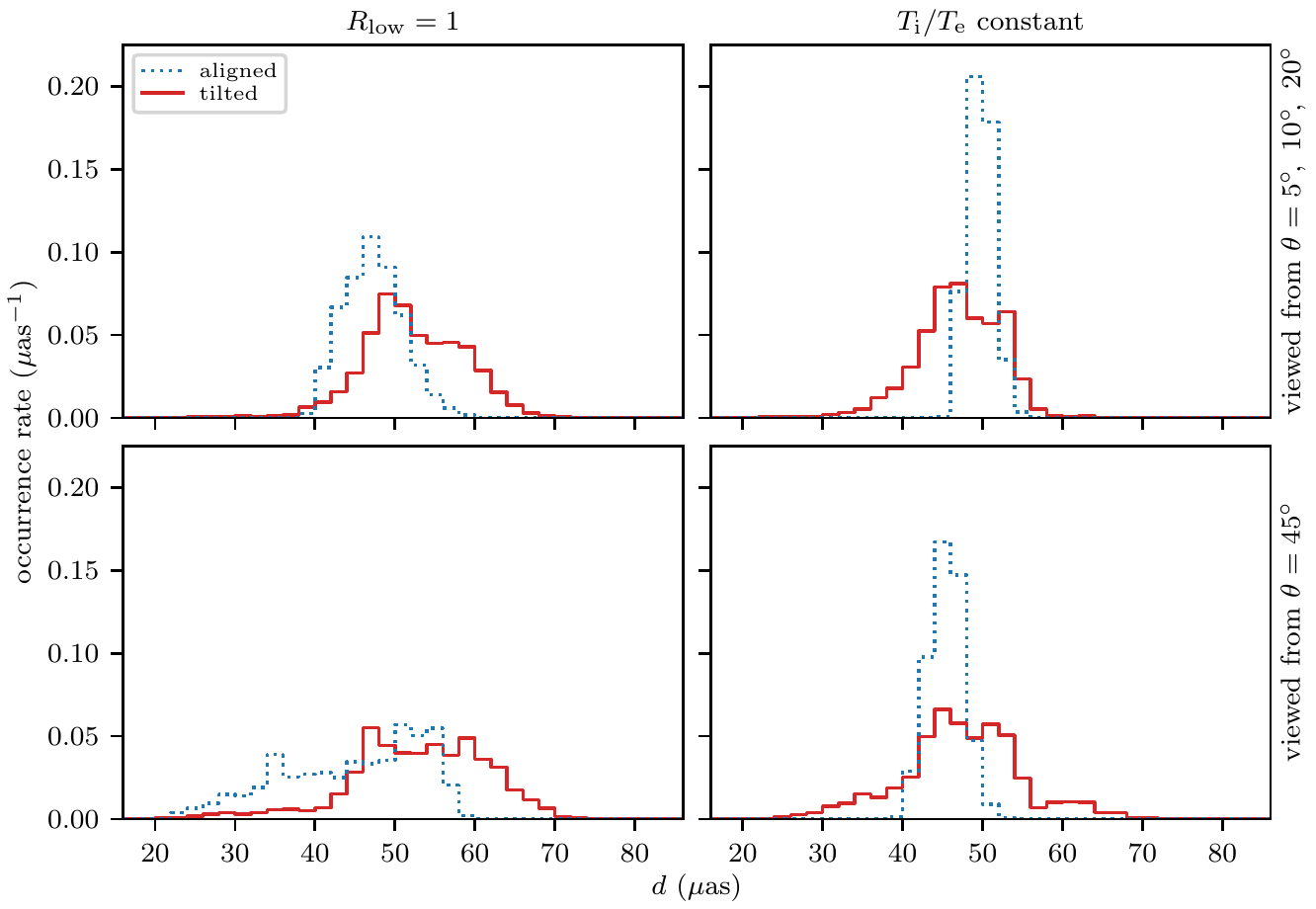}
  \caption{Distributions of measured instantaneous ring diameters at $230\ \ghz$, partitioned based on electron model and viewing angle. For all but the constant $\tti / \tte$, more face-on cases, the tilted disks have larger rings for the same black hole mass. For all but the $\rlow = 1$, least face-on cases, the tilted disk rings are more variable, even relative to their sometimes larger size. This is especially true for the constant $\tti / \tte$ models, where the aligned disk diameters are particularly constant in time. \label{fig:ring_diameter_distributions_230ghz}}
\end{figure*}

In order to better visualize these differences, we again bin the sets of measured values to construct their distributions. Here we do not subtract a linear trend, as we neither expect nor find the ring diameter to secularly change in time. These histograms are shown in Figure~\ref{fig:ring_diameter_distributions_230ghz}. Beyond the clear offset in means for all $\rlow = 1$ models, where the tilted disks have larger rings for the same black hole mass, the most striking feature in the plots is how much broader most of the distributions are for tilted disks for almost all electron models and viewing angles.

Combining all data with $\rlow = 1$, the instantaneous aligned ring diameter distribution has mean $47\ \muas$, while the tilted distribution has mean $52\ \muas$, $1.1$ times greater. For the other half of parameter space, the means are $49\ \muas$ and $47\ \muas$, respectively;\ that is, the tilted disk diameters are only $0.97$ times those of the aligned disks. Both tilt and electron model can shift these distributions significantly.

Considering next the widths of the diameter distributions, we again partition the data based on electron model and viewing inclination. Informed by Figure~\ref{fig:ring_diameter_distributions_230ghz}, we group the constant $\tti / \tte$ models together regardless of inclination, though for $\rlow = 1$ we keep the low and high inclinations separate.

Before calculating posteriors on the standard deviations of these distributions, we normalize all diameters by dividing by the corresponding diameter of the time-averaged image, thus obtaining distributions of relative ring sizes that can be easily compared to see which cases are most time variable. These time-averaged image diameters range from $45\ \muas$ to $53\ \muas$ depending on whether the disk is tilted and which portion of parameter space is being considered.

With these normalizations in place, we find the distribution of diameters relative to the diameter of the time-averaged image to have standard deviation $0.068$ for aligned disks with $\rlow = 1$ and $\theta \leq 20^\circ$, compared to $0.12$ for corresponding tilted disks. That is, the amplitude of ring size variability in time for tilted disks is $1.7$ times the value for aligned disks. The trend goes the other direction for the $\rlow = 1$, $\theta = 45^\circ$ models. The standard deviation of aligned disk diameters is $0.19$;\ for tilted disks it is $0.15$, $0.80$ times the aligned value. In the constant $\tti / \tte$ cases, the standard deviation is $0.030$ for the aligned disk and $0.13$ for the tilted disk, the latter being a factor of $4.4$ times greater.

\subsection{Structure of Bright Regions}
\label{sec:230ghz_features:bright_structure}

Even though current observations of Sgr~A* are limited to roughly $20\ \muas$ effective resolution, it can be helpful to study the structure in ray-traced images at higher resolutions to help elucidate why tilted disks are generally more variable.

\begin{figure*}
  \centering
  \includegraphics{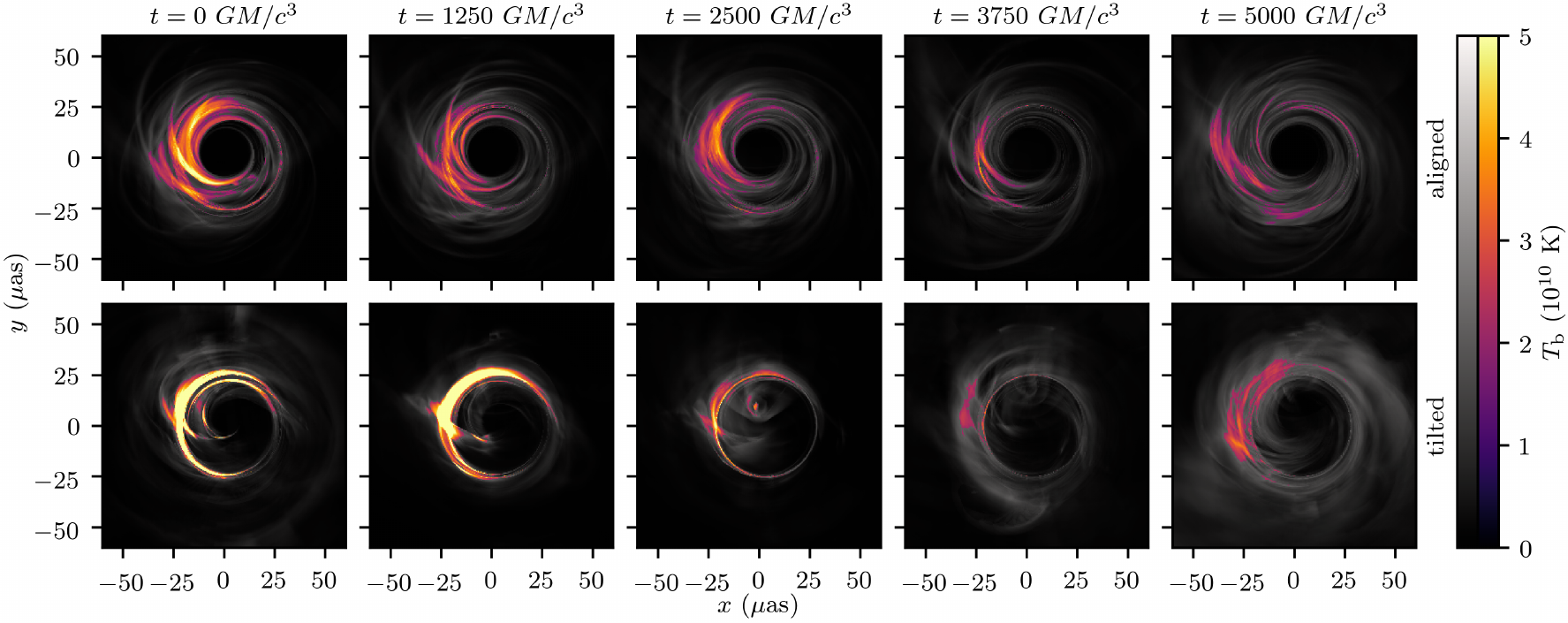}
  \caption{Snapshots at $230\ \ghz$ for the aligned and tilted disk, generated with $\rhigh = 1$, $\rlow = 40$, $\theta = 20^\circ$, and $\phi = 0^\circ$. Pixels that are colored are the brightest ones, contributing $25\%$ of the total light. The tilted disk is dominated by a few very bright, spatially and temporally coherent structures relative to the aligned disk. Snapshots made with other electron models and viewing angles show the same trend. \label{fig:images_230ghz}}
\end{figure*}

Figure~\ref{fig:images_230ghz} shows several snapshots from the $\rlow = 1$, $\rhigh = 40$, $\theta = 20^\circ$, $\phi = 0^\circ$ simulations. The images display brightness temperature (equivalently specific intensity) on a linear, perceptually uniform scale. Select pixels are colored;\ the remaining are grayscale. The colored pixels are the brightest pixels contributing to $25\%$ of the total flux (across all snapshots). In detail, we first scale the intensities in each snapshot by the factor $(2.4\ \jy + \Delta F_\nu) / F_\nu$, in order to remove the linear trend. We then find the threshold in intensity such that summing all pixels brighter than the cut results in capturing $25\%$ of the total light. These select pixels highlight which parts of the snapshot are contributing the most to the time-averaged image.

In the aligned images, there are always a number of thin, moderately bright arcs colored by our selection method. This is not unexpected for relatively face-on viewing angles. The emission is dominated by whatever turbulent structures happen to be hottest and most magnetized at a given moment. Such structures should be smaller than the whole accretion disk, they should only last for approximately a dynamical timescale, and they are likely only slightly brighter than other parts of the disk. The resulting images consist of large numbers of stochastic fluctuations, and so integrated properties seen at moderate resolutions are relatively steady in time, even as the individual bright spots move and flicker on dynamical times.

The conditions just described will not hold for all accretion flows. For example, looking nearly edge-on will result in a bright spot affixed to the approaching side of the disk. Alternatively, magnetically arrested accretion could lead to emission being dominated by a few large, temporally coherent magnetic bubbles (or their surfaces).

Tilted disks can have much the same effect owing to their out-of-plane bulk motions and standing shocks. With fewer, larger, longer-lasting structures dominating the emission, more variability in large-scale properties is to be expected on timescales between a dynamical time and an accretion time. This is what we see in Figure~\ref{fig:images_230ghz}:\ the emission is dominated by a few structures that can change position or even effectively turn off on tens to hundreds of dynamical times.

Though there is plenty of variability in both sequences of images, that of the tilted disk more readily survives blurring and time-averaging reflective of the angular and temporal resolution of EHT images. This holds for all electron models and viewing angles we consider.

\section{Image Moments at \texorpdfstring{$2.2\ \microns$}{2.2 Microns}}
\label{sec:2micron_moments}

We now shift our focus to shorter wavelengths. The images we consider here are not recalibrated;\ we simply keep the parameters used at $230\ \ghz$ and retrace the rays at a new frequency corresponding to GRAVITY observations. While GRAVITY does not have the same horizon-scale image resolution of EHT, it can precisely measure the zeroth and first moments of Sgr~A*, and so we focus on the effects that can be seen in this sort of data.

\subsection{Light Curves}
\label{sec:2micron_moments:light_curves}

\begin{figure*}
  \centering
  \includegraphics{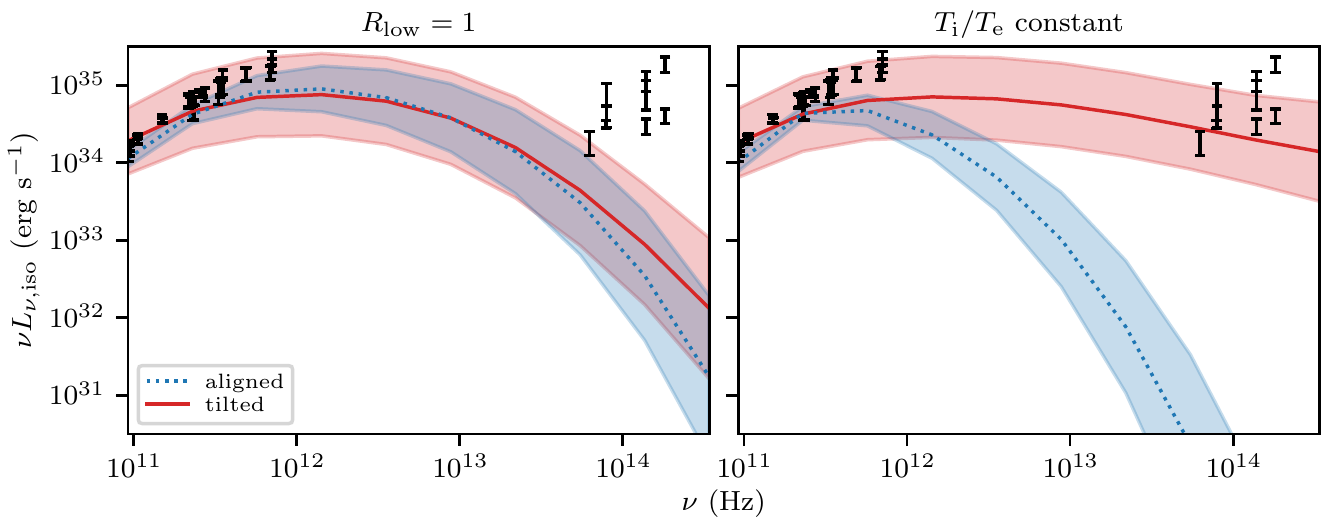}
  \caption{Spectra from the millimeter to the IR calculated from our models. At each frequency, means (lines) and standard deviations ($\pm 1$-sigma shaded) are calculated in logarithmic space from $26$ snapshots for each viewing angle and appropriate electron model. Data points show $1$-sigma ranges on observations of Sgr~A* from \citet{Falcke1998,Genzel2003,An2005,Doeleman2008,Schodel2011,Bower2015,Liu2016a,Liu2016b}. Note that in the IR these points cover the range from quiescence to flaring. Our constant $\tti / \tte$ tilted models naturally match the observed IR flux, while other models would require some extra emission at higher frequencies (e.g.\ from nonthermal electrons) to better match the data. \label{fig:spectra}}
\end{figure*}

Given that our normalization is done at $230\ \ghz$, it is not surprising that different electron models and disk configurations have vastly different IR fluxes, as shown by the spectra in Figure~\ref{fig:spectra}. The lines and shaded regions show the means and $1$-sigma ranges of spectral energies in our models, grouped according to electron model and disk orientation. The data points show the $1$-sigma ranges of observations. Even just the spectral slope between the millimeter and the IR can rule out certain models, as observed for example in \citet{Dexter2020a}. Fixing the millimeter flux to a typical value, tilted disks are systematically more luminous in the IR, especially for constant temperature ratio models. 

\begin{figure*}
  \centering
  \includegraphics{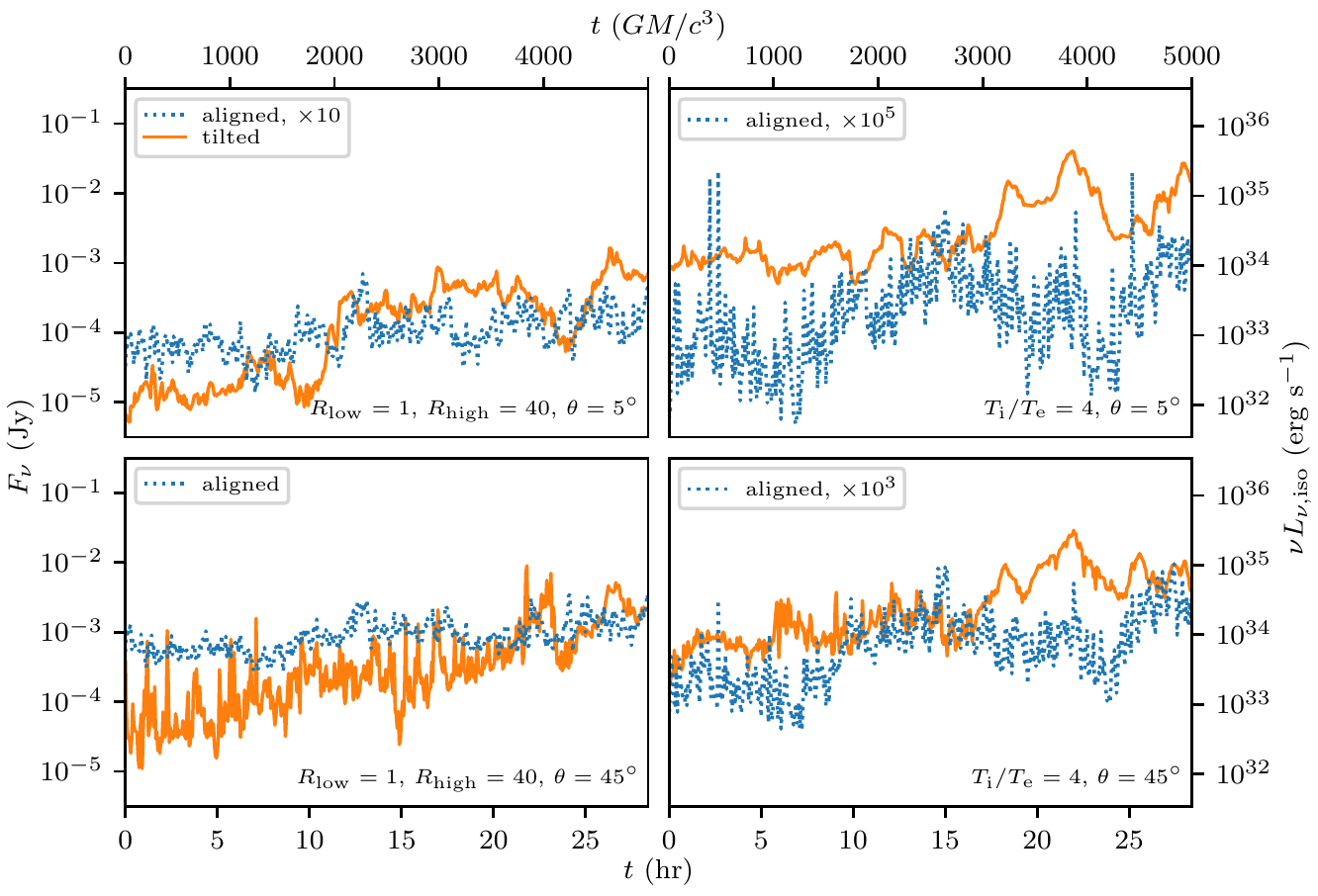}
  \caption{$2.2\ \microns$ light curves for two different electron models and viewing angles. The $\rhigh = 80$ models resemble those with $\rhigh = 40$, and light curves for $\theta = 10^\circ$ and $\theta = 20^\circ$ resemble those for $\theta = 5^\circ$. For fixed $230\ \ghz$ flux, there is usually more IR emission in tilted disks than in the corresponding aligned disks. Moreover, aligned disks display more short-term fractional variability in models with constant temperature ratios. \label{fig:light_curves_2micron}}
\end{figure*}

We proceed to examine the light curves for all of our models as they would appear in the IR at $2.2\ \microns$. Characteristic examples are shown in Figure~\ref{fig:light_curves_2micron}. Our focus here will be on the variability about the mean. We can construct the same distributions of residuals as in Section~\ref{sec:230ghz_moments:light_curves}. The IR light curves display large variability (adjacent samples can have fluxes differing by factors of close to $100$), which on a linear scale consists of very sharp spikes. We therefore work in logarithmic space, resulting in more symmetric residual distributions, which are easier to quantify.

\begin{figure*}
  \centering
  \includegraphics{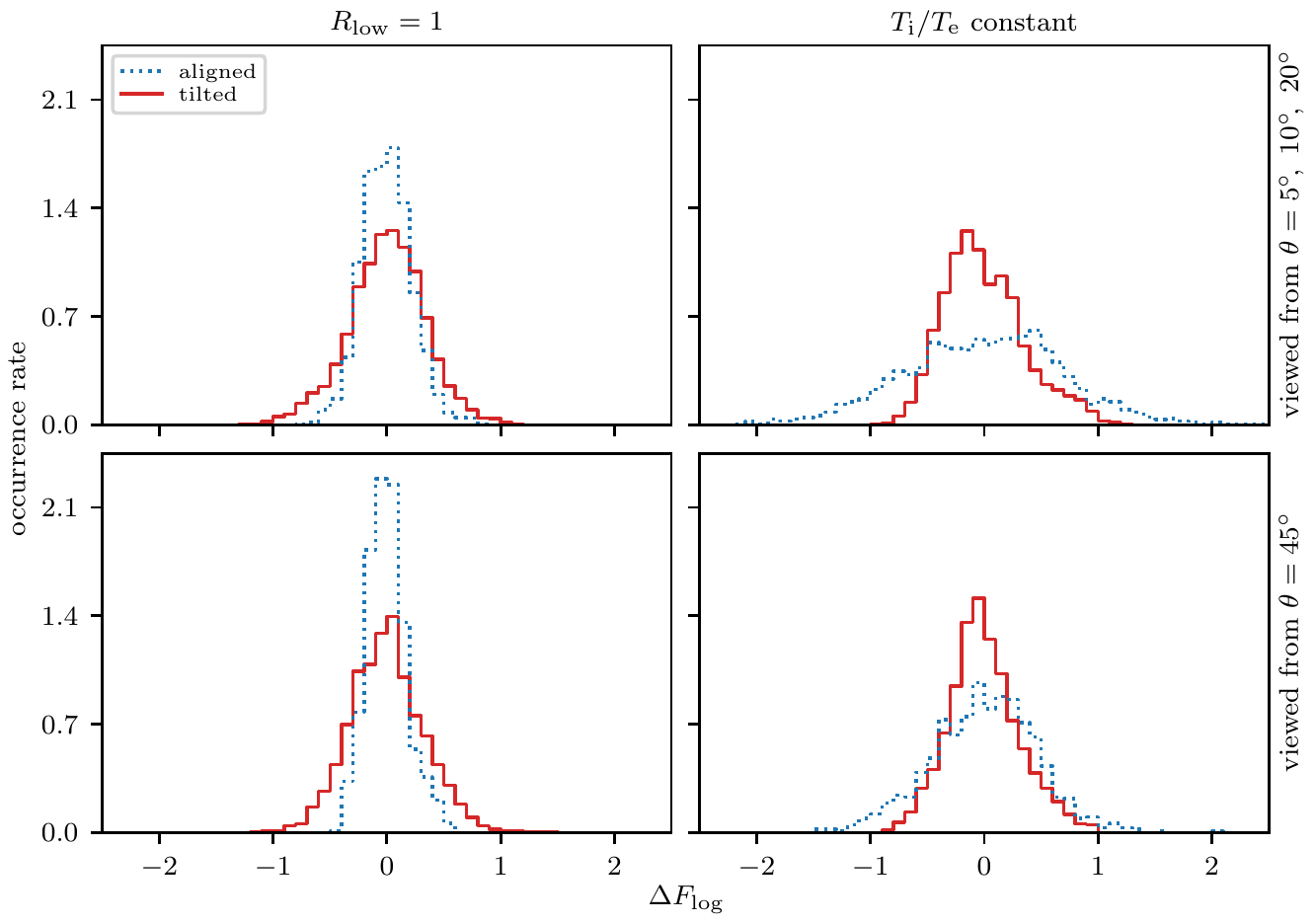}
  \caption{Distributions of logarithmic $2.2\ \microns$ light curve residuals after subtracting the linear trend, partitioned based on electron model and viewing angle. For $\rlow = 1$ electron models, tilting the disk causes the distribution to become slightly wider. However, introducing tilt results in less variability in this statistic in the electron models with constant temperature ratios, especially with nearly face-on viewing angles. \label{fig:light_curve_residuals_2micron}}
\end{figure*}

For each light curve, define
\begin{equation}
  \flog = \log_{10}\paren[\bigg]{\frac{F_\nu}{1\ \jy}}.
\end{equation}
Performing a least-squares fit of $\flog$ as a function of time to a line, we consider the set of residuals $\Delta \flog$ relative to this line. The resulting distributions are shown in Figure~\ref{fig:light_curve_residuals_2micron}.

The two $\rlow = 1$ models show a slight dependence on disk tilt in terms of these distributions. The standard deviation of the combined aligned datasets is $0.21$, whereas for the combined tilted data it is $0.33$, $60\%$ higher. In the models with constant temperature ratios, the trend is reversed. For $\theta \leq 20^\circ$, tilting the disk cases the width of the distribution to decrease $50\%$ from $0.71$ to $0.35$. The effect is reduced for $\theta = 45^\circ$:\ the distribution's width decreases $33\%$ from $0.46$ to $0.31$.

\begin{figure*}
  \centering
  \includegraphics{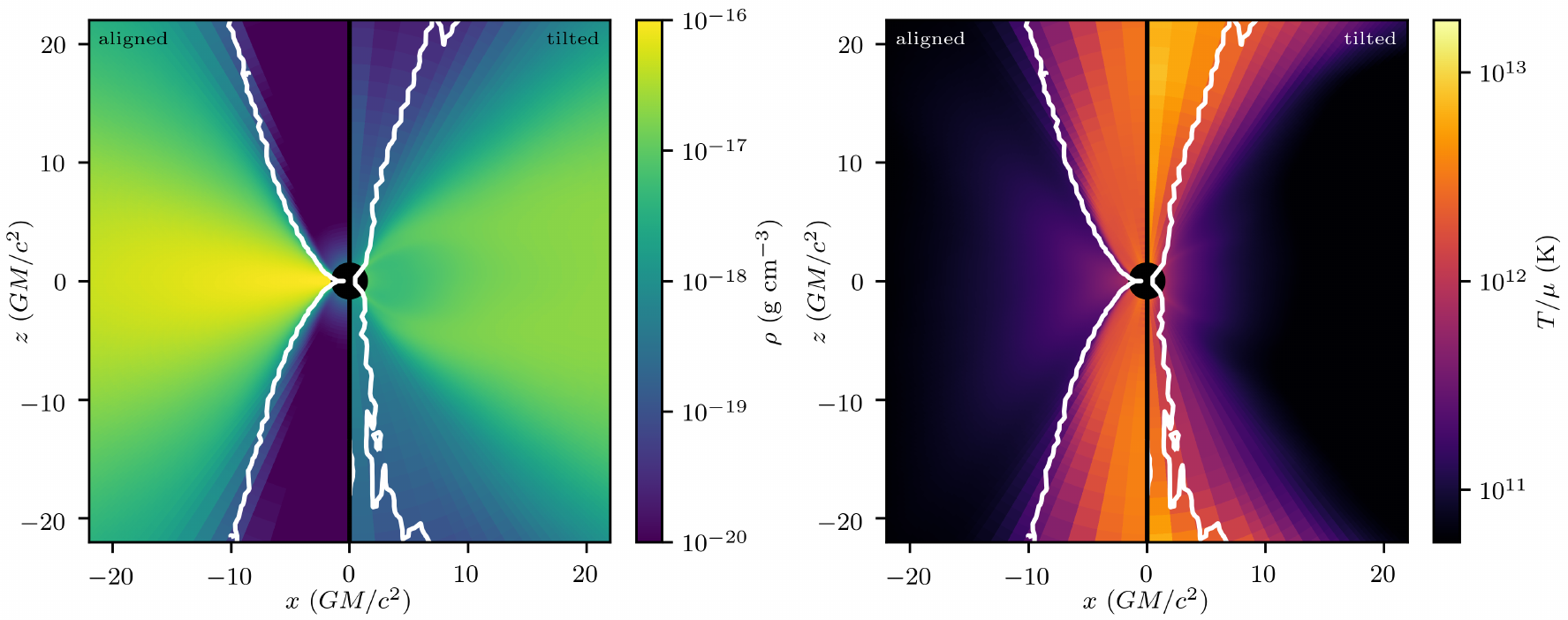}
  \caption{Density (left, normalized as appropriate for $\theta = 5^\circ$, $\rlow = 1$, $\rhigh = 40$) and total fluid temperature (right, up to a factor of the dimensionless molecular weight) in the simulations. The values are averaged in time for $6000\ G M / c^3 < t \leq 11{,}000\ G M / c^3$ and in azimuth relative to the spin axis, with the temperature density-weighted when averaging. In the tilted case, the coronal region is filled with hot material well above the density floor, contributing to emission especially in the IR. The white contours show the average location of the cutoff $\sigma = 1$. \label{fig:density_temperature}}
\end{figure*}

These results can be understood in terms of the sources of emission, i.e., the regions of high density and temperature, as shown in Figure~\ref{fig:density_temperature}. The aligned disk has a typical evacuated region along the polar axis, which is largely omitted by the $\sigma < 1$ requirement when ray tracing. This avoids contamination of the image (especially in the IR) by a hot, highly magnetized jet that has been artificially mass-loaded by numerical floors. The IR emission, therefore, is collocated with the millimeter emission, close to the origin. Tilted disks, with their more active dynamics, throw material off the disk, creating a large, hot corona that is somewhat highly magnetized but whose high density is trustworthy. Such regions are only partially excluded by the standard cut in $\sigma$. Moreover, the temperature structure of the tilted disks is not strictly monotonic;\ especially during single snapshots, the coronal region can host a hot spot considerably displaced from the black hole. In the constant temperature ratio models in particular, these regions source most of the IR emission, and they vary on longer timescales than the inner disk. In fact, much of this variability is removed by the linear fit we perform, and it would require longer simulations to more fully capture.

\begin{figure*}
  \centering
  \includegraphics{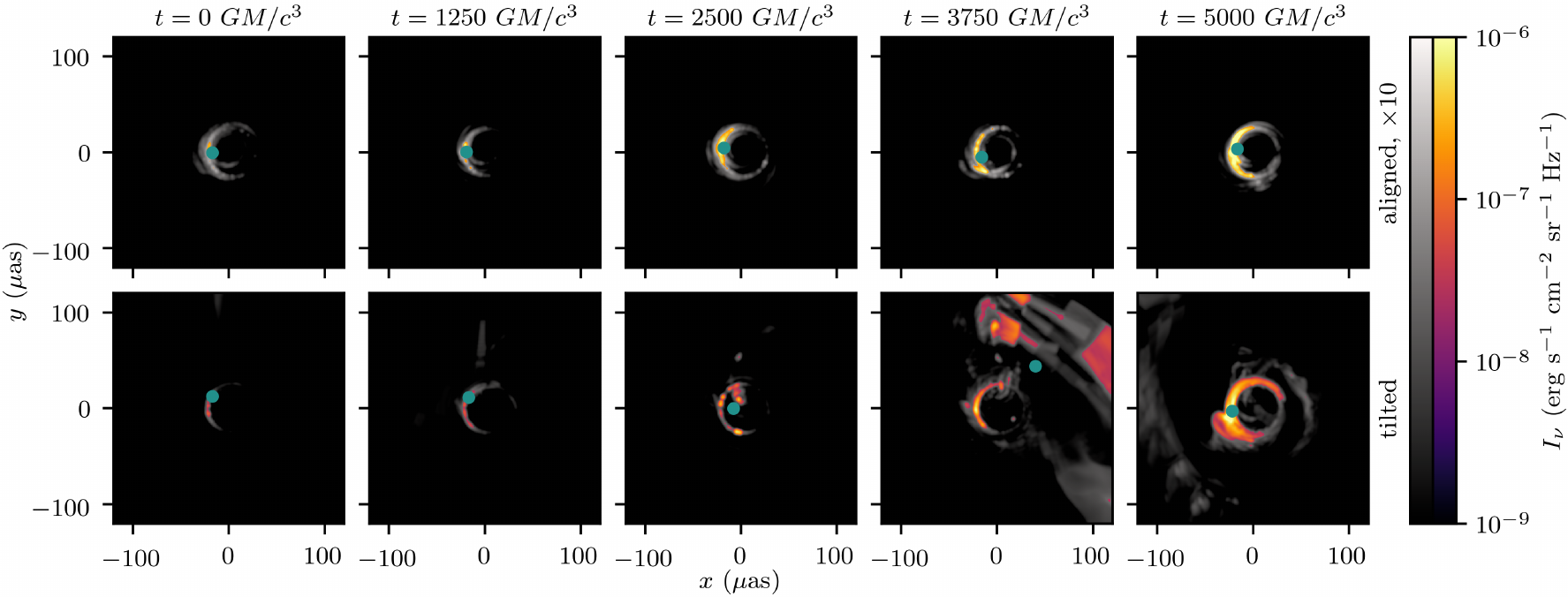}
  \caption{Snapshots at $2.2\ \microns$ for the aligned and tilted disk, generated with $\rhigh = 1$, $\rlow = 40$, $\theta = 20^\circ$, and $\phi = 0^\circ$. Pixels that are colored are the brightest ones, contributing $75\%$ of the total light. The logarithmic color scale reveals additional features, but in practice much of what is revealed contributes little to observations. At times, most of the tilted disk emission comes from the corona rather than the inner accretion disk proper. The teal points indicate the centroids of each image. \label{fig:images_2micron}}
\end{figure*}

This dichotomy between disk and coronal emission can be seen in Figure~\ref{fig:images_2micron}, the IR analogue to Figure~\ref{fig:images_230ghz}. Here we color the brightest pixels that, together, comprise $75\%$ of the total flux. In the aligned case, all of the IR emission comes from the inner accretion disk, following the behavior seen in the millimeter. The tilted accretion flow, on the other hand, occasionally emits much of its IR light from plasma outside the inner disk.

Distributions of unordered residuals characterize the amplitude of variability, but we can go further and analyze the associated timescales. It is apparent in Figure~\ref{fig:light_curves_2micron} that several of the light curves vary on timescales comparable to the sampling period ($10\ G M / c^3$), while others are much smoother.

\begin{figure*}
  \centering
  \includegraphics{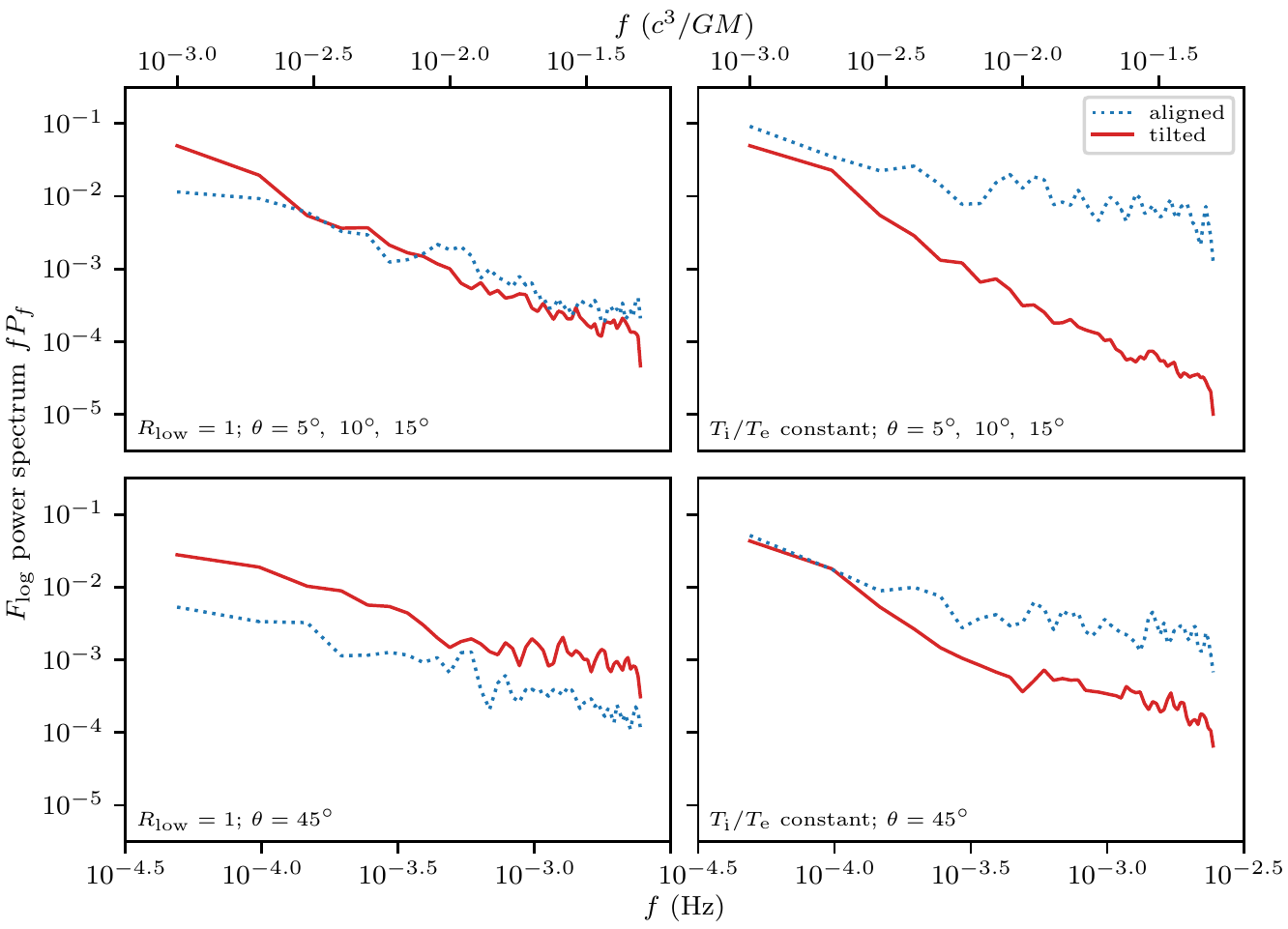}
  \caption{Power spectra for all $2.2\ \microns$ light curves, averaged together across similar electron models and viewing angles. Tilted disks display a considerable deficit of high-frequency power in logarithmic flux $\flog$ relative to aligned disks in the cases with a constant temperature ratio. Note that the aligned cases have much lower IR emission $F_\nu$. Tilt has a smaller, more ambiguous effect for $\rlow = 1$. \label{fig:power_spectra_2micron}}
\end{figure*}

This observation is made more clear by power spectra. For each light curve, we calculate the power spectrum using Welch's method on $500$ values of $\flog$. We use segments of length $100$, overlapping by $50$ points, filtered though a Hann window, though the results are insensitive to reasonable modifications of these parameters. The resulting power spectra show the same dichotomies as in other analyses:\ they depend on whether the disk is tilted, whether we are considering $\rlow = 1$ or constant $\tti / \tte$ electron models, and whether the viewing angle is far from the spin axis, but not on the details of the electron model or viewing angle. We therefore average the power spectra within each of the four evidently distinct groups. The results are shown in Figure~\ref{fig:power_spectra_2micron}.

We can see that when $\tti / \tte$ is constant, the tilted disks have much less variability in logarithmic flux relative to comparable aligned disks. There is very little difference in high-frequency power for the $\rlow = 1$ face-on models. In the case of $\rlow = 1$ at $\theta = 45^\circ$ the tilted disk IR light curves in fact have slightly more power at all frequencies.

We quantify the differences as follows. For each individual case, we measure the total (dimensionless) power in the $\flog$ light curve from $0.3\ \mhz$ to the maximum allowed by our sampling, $2.44\ \mhz$. The resulting values are grouped based on disk alignment and class of electron model (including the $\theta = 45^\circ$ case with the others). For constant $\tti / \tte$, we find the means of aligned and tilted powers to be $0.017$ and $6.3 \times 10^{-4}$, respectively;\ tilted disks have $0.038$ times the variability according to this measure. For $\rlow = 1$, the means are $0.0017$ and $0.0019$. With a ratio of central values of $1.1$, tilted disks are not distinguishable from aligned ones in this case according to this measure.

The preceding analysis of $\flog$ neglects the overall strength of IR emission. As discussed further in Section~\ref{sec:discussion}, of the cases we study, only the tilted disks with constant $\tti / \tte$ show spectra with sufficient IR flux to match observations of Sgr~A* without invoking unmodeled, nonthermal processes. It is natural to ask how close the agreement is.

\begin{figure}
  \centering
  \includegraphics{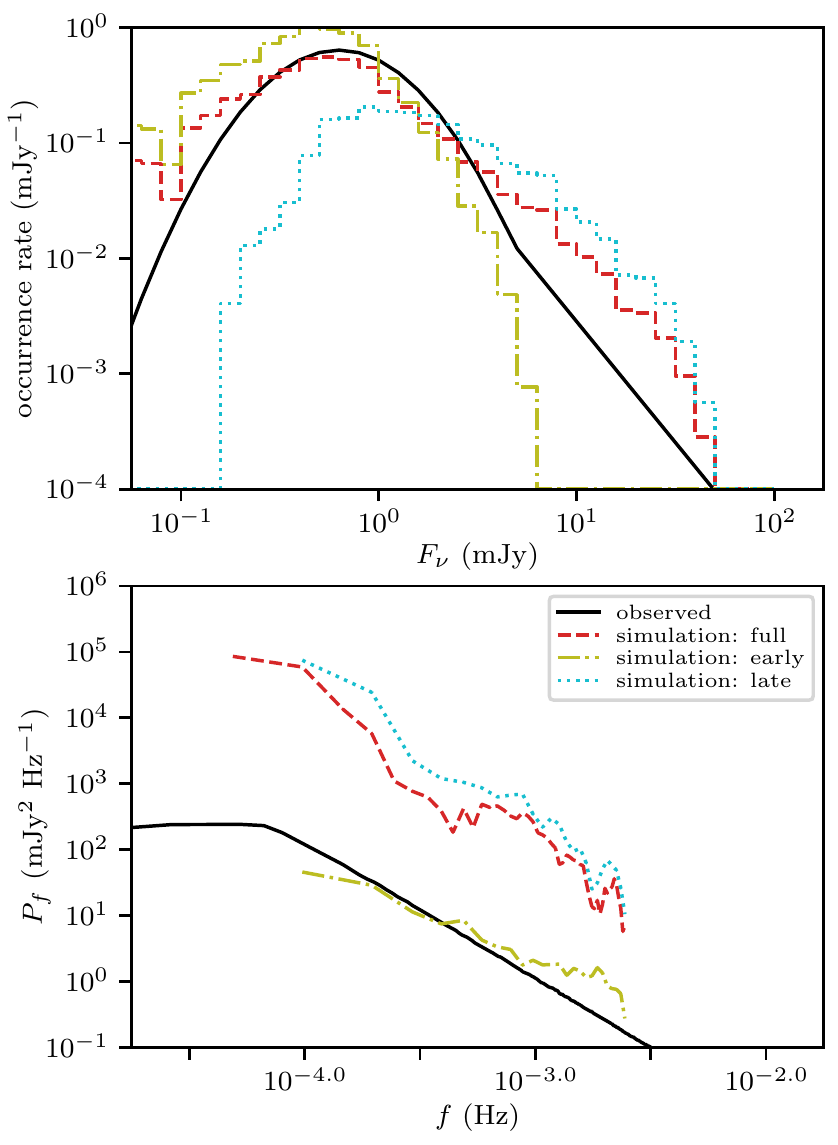}
  \caption{Comparison between observations of Sgr~A* at $2.2\ \microns$ and the corresponding values from our modeling, using the tilted disk with constant $\tti / \tte$ and nearly face-on viewing angles $\theta \leq 20^\circ$. Our light curves are both considered in their entirety and divided between the quiescent first half and flaring second half of the simulation. The distribution of fluxes from our models (top) shows an excess of time spent at particularly high fluxes (driven by the latter half of the simulation) and at particularly low fluxes (driven by the first half), relative to the observational fit from \citet{DoddsEden2011}. Our light curves are generally more variable in terms of their power spectra (bottom) than the fit from \citet{Witzel2018}, and this is due to flaring activity seen at late times in the simulation. \label{fig:statistics_2micron}}
\end{figure}

The statistical properties of the IR light curve of Sgr~A* have been quantified in a number of studies. For example, \citet{DoddsEden2011} fit the set of observed K-band ($2.2\ \microns$) fluxes to a log-normal distribution, finding that allowing for a flatter tail at larger fluxes improves the fit.\footnote{The existence of a break between quiescent and flaring states is disputed \citep{Witzel2012}. Here we are more interested in comparing to observations than in interpreting the observations themselves.} The top panel in Figure~\ref{fig:statistics_2micron} shows the fit to their 2009 dataset. We compare this to the distribution of fluxes seen in our tilted models with constant $\tti / \tte$ and with $\theta \leq 20^\circ$. Relative to observations, our models tend to spend too much time at both high and low fluxes, relative to the mode near $1\ \mjy$. Motivated by the fact that our light curves show a large, ``flaring'' event in the latter half of the simulation (see the upper right panel of Figure~\ref{fig:light_curves_2micron}), we divide our data at the halfway point in time and show the corresponding distributions. The early, ``quiescent'' state is closer to matching the observed log-normal distribution at the peak, while the remaining parts of the light curves show flares that are somewhat brighter and/or longer in duration than seen in observations.

Interestingly, the qualitative features of our tilted IR histogram roughly match those of the aligned MAD models in \citet{Chan2015b}, more so than their aligned SANE models. For example, our flux peak at approximately $1\ \mjy$ corresponds to a peak in isotropic equivalent luminosity of $10^{34}\ \erg\ \seconds^{-1}$, near where their MAD models peak. Furthermore, those MAD models tend to have excess occurrences of high IR flux above a log-normal fit to the peak, just as we see.

More recently, \citet{Witzel2018} fit for a light curve that simultaneously has a log-normal distribution of fluxes and a broken power law power spectrum. The resulting power spectrum is shown in the bottom panel of Figure~\ref{fig:statistics_2micron}.\footnote{The curve shown is not the underlying power spectrum itself, but rather the average power spectrum of $1000$ light curves generated to statistically match the best-fit parameters, as described in Appendix~B.1 of \citeauthor{Witzel2018}.} Comparing our own power spectra, we find that our models tend to be more variable on shorter timescales than Sgr~A*, and that this is driven by the latter half of the tilted simulation.

From these direct comparisons to observations, we see that our IR-bright models do not perfectly match the available data, showing more variability than is found in Sgr~A*. We note, however, that we can only simulate a relatively short light curve. Moreover, here we are only considering two electron temperature models and one disk inclination. It is reasonable to expect that smaller inclinations would produce less variability, as would electron prescriptions more akin to our $\rlow = 1$ models.

\subsection{Centroid Positions}
\label{sec:2micron_moments:centroid_positions}

The change in $2.2\ \microns$ centroid positions over time is a critical feature of GRAVITY observations of the galactic center, and so we repeat our first moment analysis at this wavelength.

\begin{figure*}
  \centering
  \includegraphics{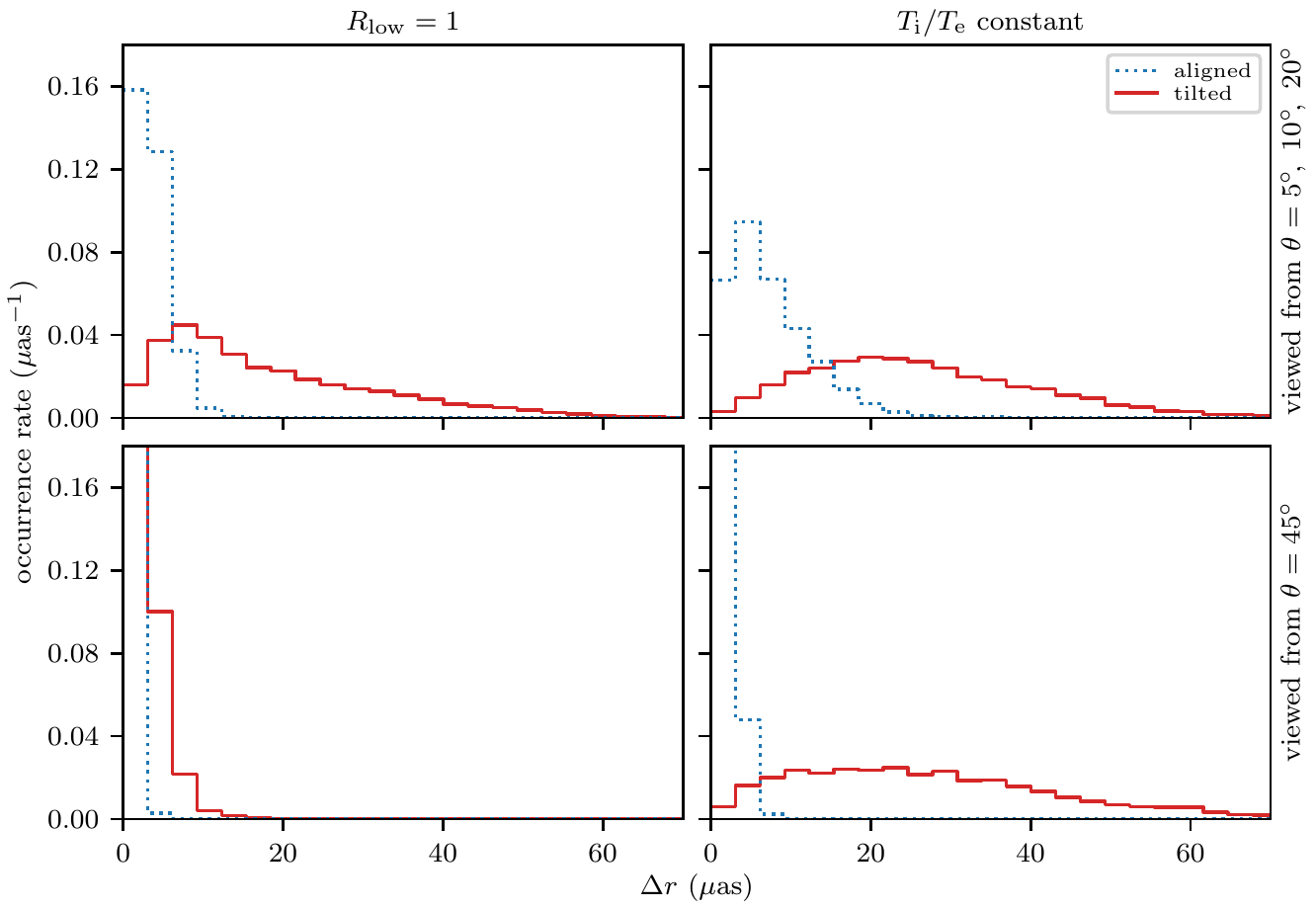}
  \caption{Distributions of displacements of instantaneous centroids from centroids of time-averaged images at $2.2\ \microns$. Results are partitioned based on the class of electron model and viewing inclination used. In all cases, the tilted disk centroids show systematically larger deviations. The axes cut off some counts of the lowest and highest $\Delta r$ values for readability. \label{fig:centroid_displacements_2micron}}
\end{figure*}

In general, our tilted disk models show large excursions of the IR centroid, of order tens of microarcseconds, as can be seen in the marked positions of the centroids in Figure~\ref{fig:images_2micron}. The differences between aligned and tilted disks can be seen clearly in histograms of displacement $\Delta r$, as shown in Figure~\ref{fig:centroid_displacements_2micron}. The centroids systematically move across more of the sky when viewing a tilted disk, regardless of which electron model and viewing angle we consider. For $\theta = 45^\circ$ the aligned images have particularly static centroids, concentrated where the plasma is moving toward the camera. Even the tilted disk shares this feature to some extent, but only for the $\rlow = 1$ electron models, and it still has larger displacements than the aligned disk.

Quantitatively, we combine all $\Delta r$ measurements except for those from cases with both $\rlow = 1$ and $\theta = 45^\circ$. The mean of the distribution for aligned disks is $5.0\ \muas$;\ for the tilted disks it is $24\ \muas$, a factor of $4.7$ larger. The omitted cases (corresponding to the bottom left panel of Figure~\ref{fig:centroid_displacements_2micron}) have means $0.98\ \muas$ and $3.0\ \muas$, respectively. Even in these cases the tilted disks are more variable by a factor of $3.1$.

As already noted in relation to the $2.2\ \microns$ light curves, tilted disks have a large amount of IR-bright plasma in their coronal regions. This leads to emission further away from the disk than can be found in aligned geometries, where only the very innermost disk has substantial emission. Moreover, the corona is influenced sufficiently by disk turbulence, reconnection events, and stochastic accretion in general to vary appreciably over the timescales considered, moving the IR centroid relatively large distances across the sky.

\begin{figure}
  \centering
  \includegraphics{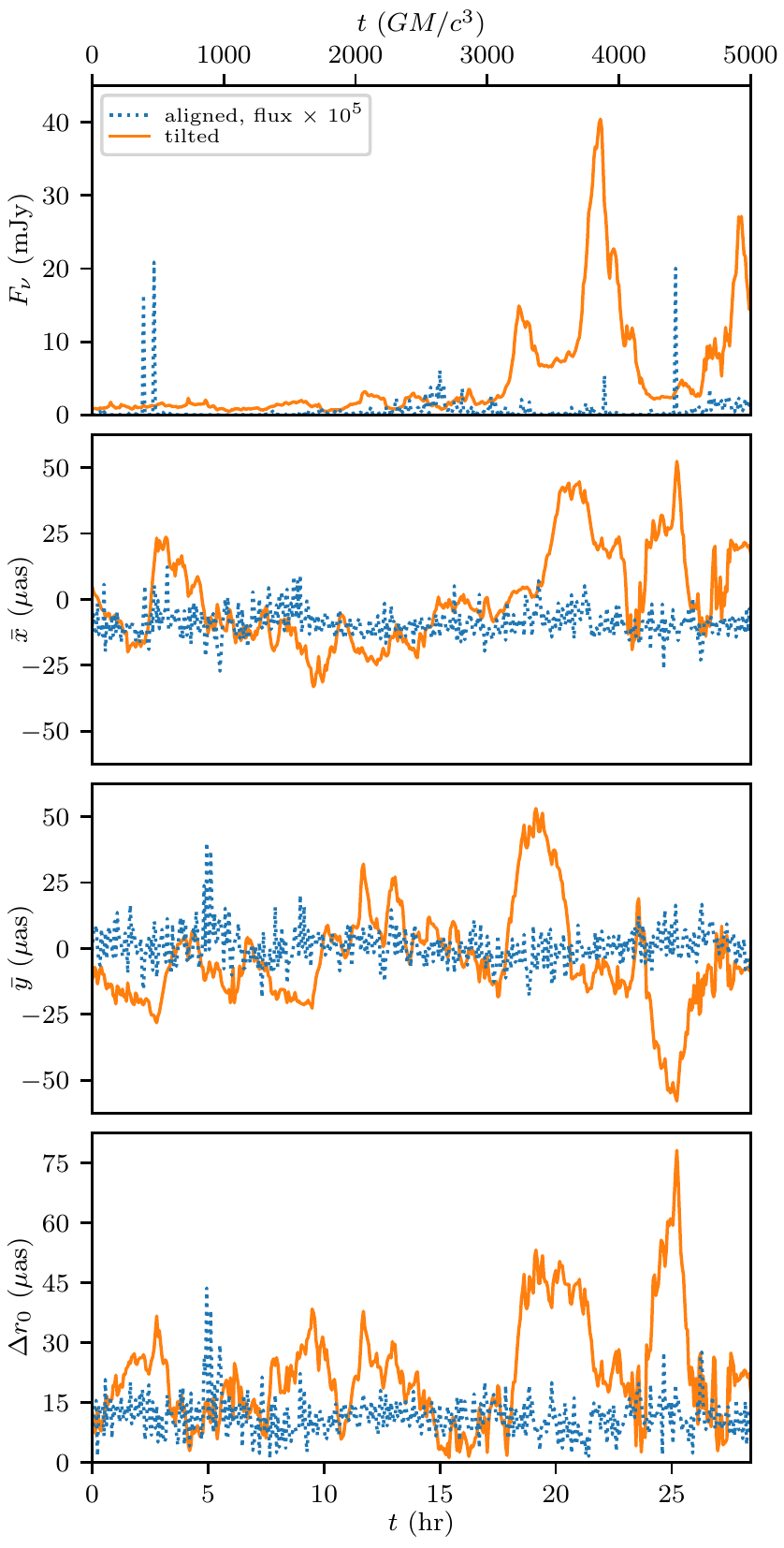}
  \caption{Instantaneous flux, centroid $x$- and $y$-positions, and displacements of instantaneous centroids from the black hole at the coordinate origin, all at $2.2\ \microns$. These curves show the results for $\tti / \tte = 4$ and $\theta = 5^\circ$. In all cases tilted disks show larger excursions, though the coherent offsets contemporaneous with a flaring event are features of constant temperature ratios and small viewing inclinations such as this. No such features are seen with aligned disks. \label{fig:centroid_positions_2micron}}
\end{figure}

This is especially notable in the constant temperature ratio models, where the latter half of the simulation shows three flaring events (perhaps better described as a double-peaked event, given the large flux between the first two peaks, followed by another single-peaked event) coincident with large centroid excursions. The light curves and centroid positions for $\tti / \tte = 4$, $\theta = 5^\circ$, and $\phi = 0^\circ$ are shown in Figure~\ref{fig:centroid_positions_2micron} for illustration. During the flare, $\xbar$ and $\ybar$ take on large values. We also show the displacement
\begin{equation}
  \Delta r_0 = \paren[\big]{\xbar^2 + \ybar^2}^{1/2}
\end{equation}
relative to the black hole, as opposed to $\Delta r$ defined earlier relative to the centroid of the averaged image. (While $\Delta r$ is akin to a direct GRAVITY observable, $\Delta r_0$ better illustrates how the flaring light tends to be at larger projected separations from the known location of the black hole in the models.)

These curves are representative of all electron models and viewing angles in the sense that all and only tilted disk images show large IR centroid displacements that are coherent on timescales of one to a few hours. The case shown in Figure~\ref{fig:centroid_positions_2micron} also displays a phase offset between $\xbar$ and $\ybar$, possibly indicating roughly circular motion. We see this with both constant $\tti / \tte$ models at all viewing angles, while the $\rlow = 1$ models do not display the same structure.

\begin{figure*}
  \centering
  \includegraphics{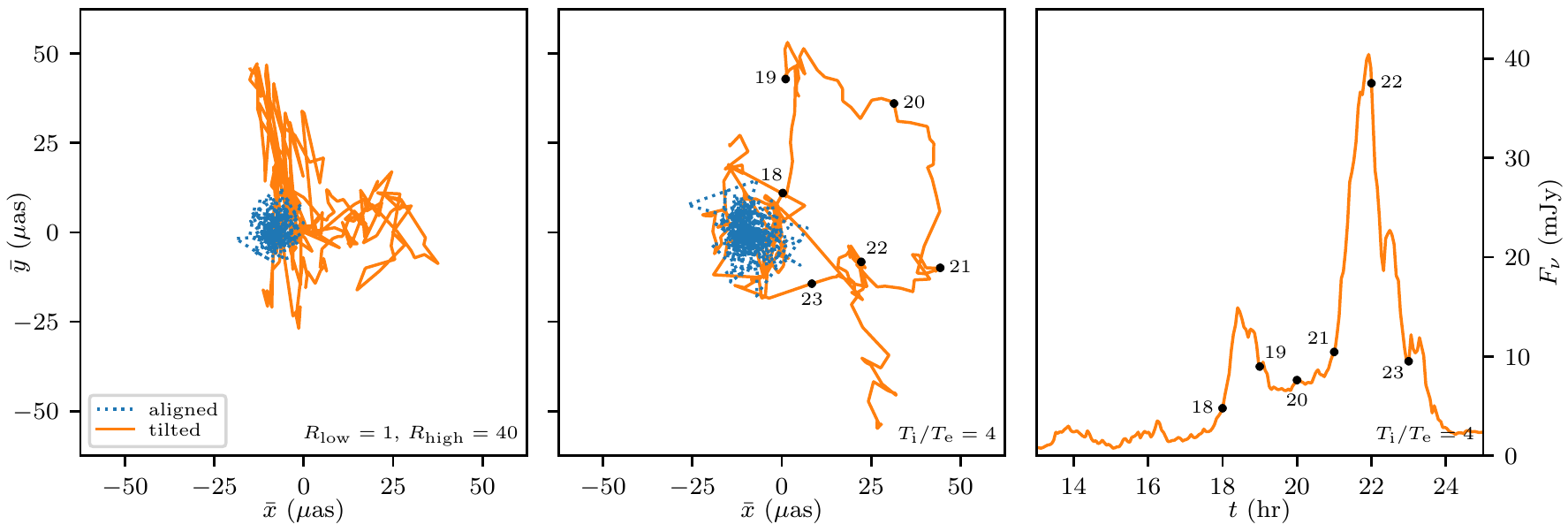}
  \caption{Paths taken by centroids of $2.2\ \microns$ light on the sky for the aligned disk and the tilted disk with two different electron models (left and center), focusing on times near the first flaring event. The paths for $\rlow = 1$, $\rhigh = 80$ are similar to those for $\rlow = 1$, $\rhigh = 40$, and likewise the constant $\tti / \tte$ models are similar to one another. Here we fix the viewing angle $\theta = 5^\circ$, $\phi = 0^\circ$, though the general trends of greater excursions in the tilted case and a circular path for constant $\tti / \tte$ hold for other angles. The right panel shows the light curve corresponding to the tilted case in the middle panel, with points labeled every hour. The circular motion corresponds to the IR flaring event seen in the latter half of the simulation. \label{fig:centroids_2micron}}
\end{figure*}

Figure~\ref{fig:centroids_2micron} shows the positions of the centroids on the sky for two representative cases, covering a timespan of $12\ \hours$. In all images we produce, the aligned disk has an IR centroid that stays relatively close to a single location, varying by only about $10\ \muas$ and by even less for $\theta = 45^\circ$. While the images are larger than this, there is simply never a time when all the emission is concentrated to one or another side of the image, as is also the case at $230\ \ghz$.

The tilted disk centroids systematically wander more in the same time interval. Both for $\rlow = 1$, $\rhigh = 40$ and for $\rlow = 1$, $\rhigh = 80$, the tilted centroid tends to have large deviations in one or two directions rather than in all directions isotropically, as pictured in the first panel. The behavior is similar among all $12$ viewing angles employed with $\theta \leq 20^\circ$. We note that for $\theta = 45^\circ$, beaming from the approaching part of the disk concentrates the light and thus centroids in a smaller region, with the centroid moving over roughly $10\ \muas$ (still more than found for the corresponding aligned disk).

As anticipated in Figure~\ref{fig:centroid_positions_2micron}, the tilted centroid in the $\tti / \tte = 4$ and $\tti / \tte = 8$ models tends to trace a roughly circular path over the course of several hours. The second panel of Figure~\ref{fig:centroids_2micron} shows one such case, where the centroid follows a clockwise path over approximately five hours, again at all viewing angles. As the camera is placed near the north polar axis of the black hole, this path is in fact retrograde. While the details of the centroid motion change with viewing angle and between the two electron models, all lines of sight with $\theta \leq 20^\circ$ show qualitatively similar behavior.

The third panel in Figure~\ref{fig:centroids_2micron} shows the light curve corresponding to this circular path. This is the same curve as shown in the upper right panel of Figure~\ref{fig:light_curves_2micron} and the top panel of Figure~\ref{fig:centroid_positions_2micron}. We label six points, spaced an hour apart, indicating how the circular excursion of the centroid coincides with the two large IR flares in the latter half of the simulation.

\section{Discussion}
\label{sec:discussion}

The material being fed into low accretion rate systems has no obvious way of aligning with the spin of the black hole, so it would be surprising to always find perfectly aligned flows at small radii. In particular this pertains to the stellar winds feeding the galactic center. At sufficiently high angular resolution, observations could directly see signatures of misaligned accretion onto a black hole, with its nonplanar shape and standing shocks. Fortunately, even with more modest resolutions the presence of tilt can still be decisively detected, albeit indirectly. We have previously argued that the size, ellipticity, and angular structure of image rings and shadows can be affected by tilt \citep{White2020b} in the same SANE simulations studied here. Here, we have extended these findings to time variability, showing how the variability of a number of image properties change when tilt is introduced into the system. This includes the time variability of the flux and image centroid, which are easier to obtain than a full image. For concreteness, we have focused our analysis on Sgr~A*, but our conclusions should apply to low accretion rate systems more broadly.

At $230\ \ghz$, tilt induces light curves with larger fluctuation amplitudes on timescales of minutes to hours for Sgr~A*, and it can even cause light curves to flicker more, changing between bright and dim with comparatively little time spent in an intermediate state. The centroid of light is displaced more, and the ring diameter in the image becomes more variable.

At $2.2\ \microns$, tilt again causes centroids to move further. While tilted IR light curves are sometimes less variable fractionally, this is a result of the aligned disks having so little flux that the light curve is dominated by flickering at the extreme tail of the electron energy distribution. In absolute terms, the tilted models have more variability at this wavelength. In fact, they can be more variable than Sgr~A*, raising the question of whether more moderate tilts can closely match observations.

In general, adding (sufficient) tilt to disks around (sufficiently rapidly) spinning black holes can induce order-unity changes to the properties of the accretion flow \citep{White2019}. One consequence of tilt is the filling of the coronal region with hotter and denser plasma than is found in the more evacuated high latitudes around aligned disks. This causes correlations in space larger than typical midplane turbulence scales. With only a small number of regions dominating the emission, the total emission's $1 / \sqrt{N}$ fractional deviations will be large, manifesting as large-amplitude fluctuations in time. Standing shocks and disk warps can contribute in the same fashion, introducing length scales into the system larger than those of turbulence in the inner parts of the disk.

Our quantitative results are summarized in Table~\ref{tab:summary}, which shows that a number of simple properties are drastically affected by disk tilt. The extent of the impact of tilt often depends on the broad class of electron temperature model employed, as well as whether the viewing angle is closely aligned with the spin axis ($\theta \leq 20^\circ$) or more edge on ($\theta = 45^\circ$). In relation to this latter dichotomy, we recall that the tilted simulation is initialized to be misaligned by $24^\circ$, and so it is not surprising that we see differences based on whether the viewing angle is more or less inclined than the disk.

\begin{deluxetable*}{lcllcC}
  \tablecaption{Summary of the effects of tilt. \label{tab:summary}}
  \tablehead{\colhead{} & \colhead{Reference} & \colhead{} & \colhead{} & \colhead{Tilted} & \colhead{Tilted/aligned} \\[-2ex]
  \colhead{Property} & \colhead{figure} & \colhead{Description} & \colhead{Dataset} & \colhead{value} & \colhead{ratio}}
  \startdata
  \sidehead{$230\ \ghz$}
  \multirow{2}{*}{\makecell[l]{Light curve \\ \tableindent variability amplitude}} & \multirow{2}{*}{\ref{fig:light_curve_residuals_230ghz}} & \multirow{2}{*}{\makecell[l]{std.\ dev.\ of \\ \tableindent $\Delta F_\nu / \ave{F_\nu}$}} & \multirow{2}{*}{all} & \multirow{2}{*}{$0.40$} & \multirow{2}{*}{2.4\phn\phn} \\
  \\
  \cmidrule(lr){4-6}
  Moment displacement & \ref{fig:centroid_displacements_230ghz} & mean of $\Delta r$ & all & $3.7\ \muas$ & 2.0\phn\phn \\
  \cmidrule(lr){4-6}
  \multirow{2}{*}{Overall ring diameter} & \multirow{2}{*}{\ref{fig:ring_diameter_distributions_230ghz}} & \multirow{2}{*}{mean of $d$} & $\rlow = 1$ & $52\ \muas$ & 1.1\phn\phn \\
  & & & $\tti / \tte$ const. & $47\ \muas$ & 0.97\phn \\
  \cmidrule(lr){4-6}
  \multirow{3}{*}{\makecell[l]{Ring diameter \\ \tableindent variability amplitude}} & \multirow{3}{*}{\ref{fig:ring_diameter_distributions_230ghz}} & \multirow{3}{*}{\makecell[l]{std.\ dev. \\ \tableindent of $d / \ave{d}$}} & $\tti / \tte$ const. & $0.13$ & 4.4\phn\phn \\
  & & & $\rlow = 1$ $\cap$ $\theta \leq 20^\circ$ & $0.12$ & 1.7\phn\phn \\
  & & & $\rlow = 1$ $\cap$ $\theta = 45^\circ$ & $0.15$ & 0.80\phn \\
  \cmidrule(lr){4-6}
  \sidehead{$2.2\ \microns$}
  \multirow{3}{*}{\makecell[l]{Light curve fractional \\ \tableindent variability amplitude}} & \multirow{3}{*}{\ref{fig:light_curve_residuals_2micron}} & \multirow{3}{*}{\makecell[l]{std.\ dev. \\ \tableindent of $\Delta \flog$}} & $\rlow = 1$ & $0.33$ & 1.6\phn\phn \\
  & & & $\tti / \tte$ const.\ $\cap$ $\theta \leq 20^\circ$ & $0.35$ & 0.50\phn \\
  & & & $\tti / \tte$ const.\ $\cap$ $\theta = 45^\circ$ & $0.31$ & 0.67\phn \\
  \cmidrule(lr){4-6}
  \multirow{2}{*}{High-frequency power} & \multirow{2}{*}{\ref{fig:power_spectra_2micron}} & \multirow{2}{*}{\makecell[l]{mean of \\ \tableindent $\int_{0.3\ \mhz}^{2.44\ \mhz} P_f \, \dd f$}} & $\rlow = 1$ & $1.9 \times 10^{-3}$ & 1.1\phn\phn \\
  & & & $\tti / \tte$ const. & $6.3 \times 10^{-4}$ & 0.038 \\
  \cmidrule(lr){4-6}
  \multirow{2}{*}{Moment displacement} & \multirow{2}{*}{\ref{fig:centroid_displacements_2micron}} & \multirow{2}{*}{mean of $\Delta r$} & $\tti / \tte$ const.\ $\cup$ $\theta \leq 20^\circ$ & $24\phd\phn\ \muas$ & 4.7\phn\phn \\
  & & & $\rlow = 1$ $\cap$ $\theta = 45^\circ$ & $\phn3.0\ \muas$ & 3.1\phn\phn \\
  \enddata
\end{deluxetable*}

There are a few caveats to this study, relating to the extent of parameter space we have covered and details in electron modeling. First, while we initialize our simulations with a poloidal magnetic field sans reversals, they do not accrete enough net vertical flux to become MAD. \Citet{Dexter2020a} found that MAD models (with reconnection-like heating, similar to constant $\tti / \tte$) tend to work best for matching Sgr~A* in terms of spectral slope and Faraday rotation. At the same time, \citet{Ressler2020b} has shown that strong magnetic fields might arise naturally from the winds feeding Sgr~A*. An investigation similar to our work here but focusing on MAD accretion is needed. Even without tilt, the different natures of SANE flows (with magnetorotational instability turbulence) and MAD flows (with magnetic Rayleigh--Taylor instability bubbles and plasmoids) lead to significantly different variability properties \citep{Chan2015b}.

The second caveat is that we only consider thermal synchrotron emission. This caveat is most relevant for the IR, where much of the light we see from Sgr~A* may be due to a high-energy, power-law tail of the electron population. Still, it is striking how IR-bright some of our tilted models are. In fact, the constant temperature ratio tilted spectra pass through the range of IR fluxes seen in Sgr~A* during quiescence and flares, as shown in Figure~\ref{fig:spectra}. The same tilted models that show circular centroid motion during flares also naturally match IR observations using just thermal electrons. At the same time, these models show more variability than Sgr~A*, as shown in Figure~\ref{fig:statistics_2micron}.

It is also possible that simply calculating electron energies in post-processing fails to capture essential features of tilted accretion flows. If electrons and ions are heated differently in one region (such as a shock) and then advected elsewhere, post-processing will struggle to assign the correct energies. To our knowledge no tilted disk with electron temperature evolved as an additional fluid quantity has been simulated;\ such a study would be particularly useful, as would further research into the plasma physics of electron heating in coronal conditions.

The simulations we use reach inflow equilibrium out to at least $r = 10\ G M / c^2$ ($50\ \muas$ for Sgr~A*);\ the polar regions reach equilibrium out to larger radii because the inflows and outflows are faster off the midplane, not being limited by the effective viscous time of the disk. The emission in both the millimeter and IR is dominated by regions comparable to or smaller than the region that is in equilibrium. That being said, for tilted disks in particular there has not been much work on what is required to reach a statistical steady state and whether the flow at smaller radii changes as inflow equilibrium is established out to larger radii. Studying this in more detail in future work would be valuable for understanding both the dynamics of tilted disks and their observational signatures.

Finally, we repeat that our findings should serve as a preliminary guide to what might be expected from tilted accretion flows in the context of present-day and near-future observations. A larger suite of simulations covering different tilts, a larger dynamic range in radii (i.e., with boundary or ad hoc initial conditions placed further from the black hole, in simulations evolved for longer times), and different magnetic field strengths would be very valuable. More investigation of edge-on accretion flows is also warranted. Still, given the nature of tilted flows, with their more complex dynamics and less evacuated coronas, we suspect that the positive correlation between tilt and variability found here will be reinforced as additional studies take into account non-planar flows.

We intend to further investigate the circular motion of the IR centroid shown in the middle panel of Figure~\ref{fig:centroids_2micron}. This motion is tantalizingly like that seen by GRAVITY during IR flaring events \citep{Gravity2018}. Modeling of hot spots \citep{Gravity2020a}, plasmoids \citep{Ball2021}, and reconnection events \citep{Dexter2020b} can produce similar centroid motions, while some three-dimensional GRMHD simulations of flux escape in MADs have some tension when trying to explain the observations \citep{Porth2021}. The large centroid motion in Figure~\ref{fig:centroids_2micron} appears for numerous viewing angles and two different electron models, but it does not occur for the aligned disk. At the moment, however, we only have the single tilted simulation, so we cannot say how often such centroid motion occurs.

With the simulations and ray tracing we have in hand, it seems clear that misalignment between gas angular momentum and black hole spin can have a significant effect on the radiation produced by accreting black holes. In particular, variability of a number of quantities increases as a disk, ceteris paribus, is tilted. This is in agreement with other investigations of tilted disks \citep{Dexter2013,Chatterjee2020}.\footnote{However, \citeauthor{Chatterjee2020} find that the motion of the brightest point in millimeter images decreases with increasing tilt, while we find that centroid motion increases. This apparent discrepancy could be due to these being different metrics. Alternatively, it could reflect differences between MAD modeling of M87 and SANE modeling of Sgr~A*.} In particular, we emphasize that the time variability of even relatively simple and coarse measurements can be strongly affected by the qualitative differences between aligned and tilted accretion. We encourage future observational data reduction and analysis to consider the possibility of misalignment, especially given that such tilt is plausibly generic in low accretion rate systems \citep{Ressler2020b}.

\acknowledgments

\strut

The authors are very grateful to Jason Dexter and Omer Blaes for discussions and suggestions that improved this work.

This research was supported in part by the National Science Foundation under grants NSF~AST~1715054 and NSF~PHY~1748958 and by a Simons Investigator award from the Simons Foundation (EQ). This work used the Extreme Science and Engineering Discovery Environment (XSEDE) cluster Stampede2 at the Texas Advanced Computing Center (TACC) through allocations AST170012 and AST200005, as well as the Princeton Research Computing cluster Tiger managed and supported by the Princeton Institute for Computational Science and Engineering (PICSciE) and the Office of Information Technology's High Performance Computing Center and Visualization Laboratory at Princeton University.

\software{\code{grtrans} \citep{Dexter2009,Dexter2016}, \code{Athena++} \citep{Stone2020,White2016}}

\bibliographystyle{aasjournal}
\bibliography{references}

\begin{thebibliography}{}
\expandafter\ifx\csname natexlab\endcsname\relax\def\natexlab#1{#1}\fi
\providecommand{\url}[1]{\href{#1}{#1}}
\providecommand{\dodoi}[1]{doi:~\href{http://doi.org/#1}{\nolinkurl{#1}}}
\providecommand{\doeprint}[1]{\href{http://ascl.net/#1}{\nolinkurl{http://ascl.net/#1}}}
\providecommand{\doarXiv}[1]{\href{https://arxiv.org/abs/#1}{\nolinkurl{https://arxiv.org/abs/#1}}}

\bibitem[{An {et~al.}(2005)An, Goss, Zhao, Hong, Roy, Rao, \& Shen}]{An2005}
An, T., Goss, W., Zhao, J.-H., {et~al.} 2005, ApJ, 634, L49,
  \dodoi{10.1086/498687}

\bibitem[{Ball {et~al.}(2021)Ball, {\"O}zel, Christian, Chan, \&
  Psaltis}]{Ball2021}
Ball, D., {\"O}zel, F., Christian, P., Chan, C.-K., \& Psaltis, D. 2021, ApJ

\bibitem[{Blandford \& Begelman(1999)}]{Blandford1999}
Blandford, R.~D., \& Begelman, M.~C. 1999, MNRAS, 303, L1,
  \dodoi{10.1046/j.1365-8711.1999.02358.x}

\bibitem[{Bower {et~al.}(2015)Bower, Markoff, Dexter, Gurwell, Moran,
  Brunthaler, Falcke, Fragile, Maitra, Marrone, Peck, Rushton, \&
  Wright}]{Bower2015}
Bower, G.~C., Markoff, S., Dexter, J., {et~al.} 2015, ApJ, 802, 69,
  \dodoi{10.1088/0004-637X/802/1/69}

\bibitem[{Bower {et~al.}(2018)Bower, Broderick, Dexter, Doeleman, Falcke, Fish,
  Johnson, Marrone, Moran, Moscibrodzka, Peck, Plambeck, \& Rao}]{Bower2018}
Bower, G.~C., Broderick, A., Dexter, J., {et~al.} 2018, ApJ, 868, 101,
  \dodoi{10.3847/1538-4357/aae983}

\bibitem[{Chael {et~al.}(2017)Chael, Narayan, \& S{\k{a}}dowski}]{Chael2017}
Chael, A.~A., Narayan, R., \& S{\k{a}}dowski, A. 2017, MNRAS, 470, 2367,
  \dodoi{10.1093/mnras/stx1345}

\bibitem[{Chan {et~al.}(2015{\natexlab{a}})Chan, Psaltis, {\"O}zel, Medeiros,
  Marrone, S{\k{a}}dowski, \& Narayan}]{Chan2015b}
Chan, C.-K., Psaltis, D., {\"O}zel, F., {et~al.} 2015{\natexlab{a}}, ApJ, 812,
  103, \dodoi{10.1088/0004-637X/812/2/103}

\bibitem[{Chan {et~al.}(2015{\natexlab{b}})Chan, Psaltis, {\"O}zel, Narayan, \&
  S{\k{a}}dowski}]{Chan2015a}
Chan, C.-K., Psaltis, D., {\"O}zel, F., Narayan, R., \& S{\k{a}}dowski, A.
  2015{\natexlab{b}}, ApJ, 799, 1, \dodoi{10.1088/0004-637X/799/1/1}

\bibitem[{Chatterjee {et~al.}(2020)Chatterjee, Younsi, Liska, Tchekhovskoy,
  Markoff, Yoon, {van~Eijnatten}, Hesp, Ingram, \&
  {van~der~Klis}}]{Chatterjee2020}
Chatterjee, K., Younsi, Z., Liska, M., {et~al.} 2020, MNRAS, 499, 362,
  \dodoi{10.1093/mnras/staa2718}

\bibitem[{Dexter(2016)}]{Dexter2016}
Dexter, J. 2016, MNRAS, 462, 115, \dodoi{10.1093/mnras/stw1526}

\bibitem[{Dexter \& Agol(2009)}]{Dexter2009}
Dexter, J., \& Agol, E. 2009, ApJ, 696, 1616,
  \dodoi{10.1088/0004-637X/696/2/1616}

\bibitem[{Dexter {et~al.}(2010)Dexter, Agol, Fragile, \& McKinney}]{Dexter2010}
Dexter, J., Agol, E., Fragile, P.~C., \& McKinney, J.~C. 2010, ApJ, 717, 1092,
  \dodoi{10.1088/0004-637X/717/2/1092}

\bibitem[{Dexter \& Fragile(2013)}]{Dexter2013}
Dexter, J., \& Fragile, P.~C. 2013, MNRAS, 432, 2252,
  \dodoi{10.1093/mnras/stt583}

\bibitem[{Dexter {et~al.}(2020{\natexlab{a}})Dexter, Jim{\'e}nez-Rosales,
  Ressler, Tchekhovskoy, Baubo{\"o}ck, {de~Zeeuw}, Eisenhauer,
  {von~Fellenberg}, Gao, Genzel, Gillessen, Habibi, Ott, Stadler, Straub, \&
  Widmann}]{Dexter2020a}
Dexter, J., Jim{\'e}nez-Rosales, A., Ressler, S., {et~al.} 2020{\natexlab{a}},
  MNRAS, 494, 4168, \dodoi{10.1093/mnras/staa922}

\bibitem[{Dexter {et~al.}(2020{\natexlab{b}})Dexter, Tchekhovskoy,
  Jim{\'e}nez-Rosales, Ressler, Baub{\"o}ck, Dallilar, {de~Zeeuw}, Eisenhauer,
  {von~Fellenberg}, Gao, Genzel, Gillessen, Habibi, Ott, Stadler, Straub, \&
  Widmann}]{Dexter2020b}
Dexter, J., Tchekhovskoy, A., Jim{\'e}nez-Rosales, A., {et~al.}
  2020{\natexlab{b}}, MNRAS, 497, 4999, \dodoi{10.1093/mnras/staa2288}

\bibitem[{Dodds-Eden {et~al.}(2011)Dodds-Eden, Gillessen, Fritz, Eisenhauer,
  Trippe, Genzel, Ott, Bartko, Pfuhl, Bower, Goldwurm, Porquet, Trap, \&
  Yusef-Zadeh}]{DoddsEden2011}
Dodds-Eden, K., Gillessen, S., Fritz, T., {et~al.} 2011, ApJ, 728, 37,
  \dodoi{10.1088/0004-637X/728/1/37}

\bibitem[{Doeleman {et~al.}(2008)Doeleman, Weintroub, Rogers, Plambeck, Freund,
  Tilanus, Friberg, Ziurys, Moran, Corey, Young, Smythe, Titus, Marrone,
  Cappallo, Bock, Bower, Chamberlin, Davis, Krichbaum, Lamb, Maness, Niell,
  Roy, Strittmatter, Werthimer, Whitney, \& Woody}]{Doeleman2008}
Doeleman, S.~S., Weintroub, J., Rogers, A.~E., {et~al.} 2008, Natur, 455, 78,
  \dodoi{10.1038/nature07245}

\bibitem[{Dolence {et~al.}(2012)Dolence, Gammie, Shiokawa, \&
  Noble}]{Dolence2012}
Dolence, J.~C., Gammie, C.~F., Shiokawa, H., \& Noble, S.~C. 2012, ApJL, 746,
  L10, \dodoi{10.1088/2041-8205/746/1/L10}

\bibitem[{Falcke {et~al.}(1998)Falcke, Goss, Matsuo, Teuben, Zhao, \&
  Zylka}]{Falcke1998}
Falcke, H., Goss, W., Matsuo, H., {et~al.} 1998, ApJ, 499, 731,
  \dodoi{10.1086/305687}

\bibitem[{Fishbone \& Moncrief(1976)}]{Fishbone1976}
Fishbone, L.~G., \& Moncrief, V. 1976, ApJ, 207, 962, \dodoi{10.1086/154565}

\bibitem[{Foucart {et~al.}(2017)Foucart, Chandra, Gammie, Quataert, \&
  Tchekhovskoy}]{Foucart2017}
Foucart, F., Chandra, M., Gammie, C.~F., Quataert, E., \& Tchekhovskoy, A.
  2017, MNRAS, 470, 2240, \dodoi{10.1093/mnras/stx1368}

\bibitem[{Fragile \& Anninos(2005)}]{Fragile2005}
Fragile, P.~C., \& Anninos, P. 2005, ApJ, 623, 347, \dodoi{10.1086/428433}

\bibitem[{Fragile \& Blaes(2008)}]{Fragile2008}
Fragile, P.~C., \& Blaes, O.~M. 2008, ApJ, 687, 757, \dodoi{10.1086/591936}

\bibitem[{Fragile {et~al.}(2007)Fragile, Blaes, Anninos, \&
  Salmonson}]{Fragile2007}
Fragile, P.~C., Blaes, O.~M., Anninos, P., \& Salmonson, J.~D. 2007, ApJ, 668,
  417, \dodoi{10.1086/521092}

\bibitem[{Gebhardt {et~al.}(2011)Gebhardt, Adams, Richstone, Lauer, Faber,
  G{\"u}ltekin, Murphy, \& Tremaine}]{Gebhardt2011}
Gebhardt, K., Adams, J., Richstone, D., {et~al.} 2011, ApJ, 729, 119,
  \dodoi{10.1088/0004-637X/729/2/119}

\bibitem[{Generozov {et~al.}(2014)Generozov, Blaes, Fragile, \&
  Henisey}]{Generozov2014}
Generozov, A., Blaes, O., Fragile, P.~C., \& Henisey, K.~B. 2014, ApJ, 780, 81,
  \dodoi{10.1088/0004-637X/780/1/81}

\bibitem[{Genzel {et~al.}(2003)Genzel, Sch{\"o}del, Ott, Eckart, Alexander,
  Lacombe, Rouan, \& Aschenbach}]{Genzel2003}
Genzel, R., Sch{\"o}del, R., Ott, T., {et~al.} 2003, Natur, 425, 934,
  \dodoi{10.1038/nature02065}

\bibitem[{Ghez {et~al.}(2003)Ghez, Duch{\^e}ne, Matthews, Hornstein, Tanner,
  Larkin, Morris, Becklin, Salim, Kremenek, Thompson, Soifer, Neugebauer, \&
  McLean}]{Ghez2003}
Ghez, A., Duch{\^e}ne, G., Matthews, K., {et~al.} 2003, ApJ, 586, L127,
  \dodoi{10.1086/374804}

\bibitem[{Ghez {et~al.}(2008)Ghez, Salim, Weinberg, Lu, Do, Dunn, Matthews,
  Morris, Yelda, Becklin, Kremenek, Milosavljevic, \& Naiman}]{Ghez2008}
Ghez, A., Salim, S., Weinberg, N., {et~al.} 2008, ApJ, 689, 1044,
  \dodoi{10.1086/592738}

\bibitem[{Gillessen {et~al.}(2009)Gillessen, Eisenhauer, Trippe, Alexander,
  Genzel, Martins, \& Ott}]{Gillessen2009}
Gillessen, S., Eisenhauer, F., Trippe, S., {et~al.} 2009, ApJ, 692, 1075,
  \dodoi{10.1088/0004-637X/692/2/1075}

\bibitem[{Gillessen {et~al.}(2019)Gillessen, Plewa, Widmann, {von~Fellenberg},
  Schartmann, Habibi, {Jimenez~Rosales}, Baub{\"o}ck, Dexter, Gao, Waisberg,
  Eisenhauer, Pfuhl, Ott, Burkert, {de~Zeeuw}, \& Genzel}]{Gillessen2019}
Gillessen, S., Plewa, P., Widmann, F., {et~al.} 2019, ApJ, 871, 126,
  \dodoi{10.3847/1538-4357/aaf4f8}

\bibitem[{{GRAVITY Collaboration}(2018)}]{Gravity2018}
{GRAVITY Collaboration}. 2018, A\&A, 618, L10,
  \dodoi{10.1051/0004-6361/201834294}

\bibitem[{{GRAVITY Collaboration}(2020{\natexlab{a}})}]{Gravity2020b}
---. 2020{\natexlab{a}}, A\&A, 636, L5, \dodoi{10.1051/0004-6361/202037813}

\bibitem[{{GRAVITY Collaboration}(2020{\natexlab{b}})}]{Gravity2020a}
---. 2020{\natexlab{b}}, A\&A, 635, A143, \dodoi{10.1051/0004-6361/201937233}

\bibitem[{Howes(2010)}]{Howes2010}
Howes, G. 2010, MNRAS, 409, L104, \dodoi{10.1111/j.1745-3933.2010.00958.x}

\bibitem[{Kawazura {et~al.}(2019)Kawazura, Barnes, \&
  Schekochihin}]{Kawazura2019}
Kawazura, Y., Barnes, M., \& Schekochihin, A.~A. 2019, PNAS, 116, 771,
  \dodoi{10.1073/pnas.1812491116}

\bibitem[{Liska {et~al.}(2018)Liska, Hesp, Tchekhovskoy, Ingram,
  {van~der~Klis}, \& Markoff}]{Liska2018}
Liska, M., Hesp, C., Tchekhovskoy, A., {et~al.} 2018, MNRAS, 474, L81,
  \dodoi{10.1093/mnrasl/slx174}

\bibitem[{Liska {et~al.}(2020)Liska, Tchekhovskoy, \& Quataert}]{Liska2020}
Liska, M., Tchekhovskoy, A., \& Quataert, E. 2020, MNRAS, 494, 3656,
  \dodoi{10.1093/mnras/staa955}

\bibitem[{Liu {et~al.}(2016{\natexlab{a}})Liu, Wright, Zhao, Mills,
  Requena-Torres, Matsushita, Mart{\'i}n, Ott, Morris, Longmore, Brinkerink, \&
  Falcke}]{Liu2016a}
Liu, H.~B., Wright, M.~C., Zhao, J.-H., {et~al.} 2016{\natexlab{a}}, A\&A, 593,
  A44, \dodoi{10.1051/0004-6361/201628176}

\bibitem[{Liu {et~al.}(2016{\natexlab{b}})Liu, Wright, Zhao, Brinkerink, Ho,
  Mills, Mart{\'i}n, Falcke, Matsushita, \& Mart{\'i}-Vidal}]{Liu2016b}
---. 2016{\natexlab{b}}, A\&A, 593, A107, \dodoi{10.1051/0004-6361/201628731}

\bibitem[{Lu {et~al.}(2009)Lu, Ghez, Hornstein, Morris, Becklin, \&
  Matthews}]{Lu2009}
Lu, J., Ghez, A., Hornstein, S., {et~al.} 2009, ApJ, 690, 1463,
  \dodoi{10.1088/0004-637/690/2/1463}

\bibitem[{Marrone {et~al.}(2007)Marrone, Moran, Zhao, \& Rao}]{Marrone2007}
Marrone, D.~P., Moran, J.~M., Zhao, J.-H., \& Rao, R. 2007, ApJ, 654, L57,
  \dodoi{10.1086/510850}

\bibitem[{Matusmoto {et~al.}(2020)Matusmoto, Chan, \& Piran}]{Matsumoto2020}
Matusmoto, T., Chan, C.-H., \& Piran, T. 2020, MNRAS, 497, 2385,
  \dodoi{10.1093/mnras/staa2095}

\bibitem[{McKinney {et~al.}(2012)McKinney, Tchekhovskoy, \&
  Blandford}]{McKinney2012}
McKinney, J.~C., Tchekhovskoy, A., \& Blandford, R.~D. 2012, MNRAS, 423, 3083,
  \dodoi{10.1111/j.1365-2966.2012.21074.x}

\bibitem[{Medeiros {et~al.}(2017)Medeiros, kwan Chan, {\"O}zel, Psaltis, Kim,
  Marrone, \& S{\k{a}}dowski}]{Medeiros2017}
Medeiros, L., kwan Chan, C., {\"O}zel, F., {et~al.} 2017, ApJ, 844, 35,
  \dodoi{10.3847/1538-4357/aa7751}

\bibitem[{Medeiros {et~al.}(2018)Medeiros, kwan Chan, {\"O}zel, Psaltis, Kim,
  Marrone, \& S{\k{a}}dowski}]{Medeiros2018}
---. 2018, ApJ, 856, 163, \dodoi{10.3847/1538-4357/aab204}

\bibitem[{Mo{\'s}cibrodzka {et~al.}(2016)Mo{\'s}cibrodzka, Falcke, \&
  Shiokawa}]{Moscibrodzka2016}
Mo{\'s}cibrodzka, M., Falcke, H., \& Shiokawa, H. 2016, A\&A, 586, A38,
  \dodoi{10.1051/0004-6361/201526630}

\bibitem[{Mo{\'s}cibrodzka {et~al.}(2011)Mo{\'s}cibrodzka, Gammie, Dolence, \&
  Shiokawa}]{Moscibrodzka2011}
Mo{\'s}cibrodzka, M., Gammie, C., Dolence, J., \& Shiokawa, H. 2011, ApJ, 735,
  9, \dodoi{10.1088/0004-637X/735/1/9a}

\bibitem[{Narayan {et~al.}(2003)Narayan, Igumenshchev, \&
  Abramowicz}]{Narayan2003}
Narayan, R., Igumenshchev, I.~V., \& Abramowicz, M.~A. 2003, PASJ, 55, L69,
  \dodoi{10.1093/pasj/55.6.L69}

\bibitem[{Narayan {et~al.}(1998)Narayan, Mahadevan, Grindlay, Popham, \&
  Gammie}]{Narayan1998}
Narayan, R., Mahadevan, R., Grindlay, J.~E., Popham, R.~G., \& Gammie, C. 1998,
  ApJ, 492, 554, \dodoi{10.1086/305070}

\bibitem[{Narayan {et~al.}(2012)Narayan, S{\k{a}}dowski, Penna, \&
  Kulkarni}]{Narayan2012}
Narayan, R., S{\k{a}}dowski, A., Penna, R.~F., \& Kulkarni, A.~K. 2012, MNRAS,
  426, 3241, \dodoi{10.1111/j.1365-2966.2012.22002.x}

\bibitem[{Oliphant(2006)}]{Oliphant2006}
Oliphant, T.~E. 2006, in All Faculty Publications, BYU ScholarsArchive, 278

\bibitem[{Paumard {et~al.}(2006)Paumard, Genzel, Martins, Nayakshin,
  Beloborodov, Levin, Trippe, Eisenhauer, Ott, Gillessen, Abuter, Cuadra,
  Alexander, \& Sternberg}]{Paumard2006}
Paumard, T., Genzel, R., Martins, F., {et~al.} 2006, ApJ, 643, 1011,
  \dodoi{10.1086/503273}

\bibitem[{Porth {et~al.}(2021)Porth, Mizuno, Younsi, \& Fromm}]{Porth2021}
Porth, O., Mizuno, Y., Younsi, Z., \& Fromm, C. 2021, MNRAS, 502, 2023,
  \dodoi{10.1093/mnras/stab163}

\bibitem[{Porth {et~al.}(2019)Porth, Chatterjee, Narayan, Gammie, Mizuno,
  Anninos, Baker, Bugli, kwan Chan, Davelaar, {Del~Zanna}, Etienne, Fragile,
  Kelly, Liska, Markoff, McKinney, Mishra, Noble, Olivares, Prather, Rezzolla,
  Ryan, Stone, Tomei, White, Younsi, \& {The Event Horizon Telescope
  Collaboration}}]{Porth2019}
Porth, O., Chatterjee, K., Narayan, R., {et~al.} 2019, ApJS, 243, 26,
  \dodoi{10.3847/1538-4365/ab29fd}

\bibitem[{Punsly {et~al.}(2009)Punsly, Igumenshchev, \& Hirose}]{Punsly2009}
Punsly, B., Igumenshchev, I.~V., \& Hirose, S. 2009, ApJ, 704, 1065,
  \dodoi{10.1088/0004-637X/704/2/1065}

\bibitem[{Ressler {et~al.}(2020{\natexlab{a}})Ressler, Quataert, \&
  Stone}]{Ressler2020a}
Ressler, S., Quataert, E., \& Stone, J. 2020{\natexlab{a}}, MNRAS, 492, 3272,
  \dodoi{10.1093/mnras/stz3605}

\bibitem[{Ressler {et~al.}(2015)Ressler, Tchekhovskoy, Quataert, Chandra, \&
  Gammie}]{Ressler2015}
Ressler, S., Tchekhovskoy, A., Quataert, E., Chandra, M., \& Gammie, C. 2015,
  MNRAS, 454, 1848, \dodoi{10.1093/mnras/stv2084}

\bibitem[{Ressler {et~al.}(2020{\natexlab{b}})Ressler, White, Quataert, \&
  Stone}]{Ressler2020b}
Ressler, S.~M., White, C.~J., Quataert, E., \& Stone, J.~M. 2020{\natexlab{b}},
  ApJL, 896, L6, \dodoi{10.3847/2041-8213/ab9532}

\bibitem[{Rowan {et~al.}(2017)Rowan, Sironi, \& Narayan}]{Rowan2017}
Rowan, M.~E., Sironi, L., \& Narayan, R. 2017, ApJ, 850, 29,
  \dodoi{10.3847/1538-4357/aa9380}

\bibitem[{Ryan {et~al.}(2018)Ryan, Ressler, Dolence, Gammie, \&
  Quataert}]{Ryan2018}
Ryan, B.~R., Ressler, S.~M., Dolence, J.~C., Gammie, C., \& Quataert, E. 2018,
  ApJ, 864, 126, \dodoi{10.3847/1538-4357/aad73a}

\bibitem[{S{\k{a}}dowski {et~al.}(2013)S{\k{a}}dowski, Narayan, Penna, \&
  Zhu}]{Sadowski2013}
S{\k{a}}dowski, A., Narayan, R., Penna, R., \& Zhu, Y. 2013, MNRAS, 436, 3856,
  \dodoi{10.1093/mnras/stt1881}

\bibitem[{S{\k{a}}dowski {et~al.}(2017)S{\k{a}}dowski, Wielgus, Narayan,
  Abarca, McKinney, \& Chael}]{Sadowski2017}
S{\k{a}}dowski, A., Wielgus, M., Narayan, R., {et~al.} 2017, MNRAS, 466, 705,
  \dodoi{10.1093/mnras/stw3116}

\bibitem[{Sch{\"o}del {et~al.}(2011)Sch{\"o}del, Morris, Muzic, Alberdi, Meyer,
  Eckart, \& Gezari}]{Schodel2011}
Sch{\"o}del, R., Morris, M., Muzic, K., {et~al.} 2011, A\&A, 532, A83,
  \dodoi{10.1051/0004-6361/201116994}

\bibitem[{Stone {et~al.}(2020)Stone, Tomida, White, \& Felker}]{Stone2020}
Stone, J.~M., Tomida, K., White, C.~J., \& Felker, K.~G. 2020, ApJS, 249, 4,
  \dodoi{10.3847/1538-4365/ab929b}

\bibitem[{Tchekhovskoy {et~al.}(2011)Tchekhovskoy, Narayan, \&
  McKinney}]{Tchekhovskoy2011}
Tchekhovskoy, A., Narayan, R., \& McKinney, J.~C. 2011, MNRAS, 418, L79,
  \dodoi{10.1111/j.1745-3933.2011.01147.x}

\bibitem[{{The Event Horizon Telescope
  Collaboration}(2019{\natexlab{a}})}]{EHT2019a}
{The Event Horizon Telescope Collaboration}. 2019{\natexlab{a}}, ApJL, 875, L1,
  \dodoi{10.3847/2041-8213/ab0ec7}

\bibitem[{{The Event Horizon Telescope
  Collaboration}(2019{\natexlab{b}})}]{EHT2019f}
---. 2019{\natexlab{b}}, ApJL, 875, L6, \dodoi{10.3847/2041-8213/ab1141}

\bibitem[{{The Event Horizon Telescope
  Collaboration}(2019{\natexlab{c}})}]{EHT2019b}
---. 2019{\natexlab{c}}, ApJL, 875, L2, \dodoi{10.3847/2041-8213/ab0c96}

\bibitem[{{The GRAVITY Collaboration}(2019)}]{Gravity2019}
{The GRAVITY Collaboration}. 2019, A\&A, 625, L10,
  \dodoi{10.1051/0004-6361/201935656}

\bibitem[{Walsh {et~al.}(2013)Walsh, Barth, Ho, \& Sarzi}]{Walsh2013}
Walsh, J.~L., Barth, A.~J., Ho, L.~C., \& Sarzi, M. 2013, ApJ, 770, 86,
  \dodoi{10.1088/0004-637X/770/2/86}

\bibitem[{Wang {et~al.}(2013)Wang, Nowak, Markoff, Baganoff, Nayakshin, Yuan,
  Cuadra, Davis, Dexter, Fabian, Grosso, Haggard, Houck, Ji, Li, Neilsen,
  Porquet, Ripple, \& Shcherbakov}]{Wang2013}
Wang, Q., Nowak, M., Markoff, S., {et~al.} 2013, Sci, 341, 981,
  \dodoi{10.1126/science.1240755}

\bibitem[{Werner {et~al.}(2018)Werner, Uzdensky, Begelman, Cerutti, \&
  Nalewajko}]{Werner2018}
Werner, G., Uzdensky, D., Begelman, M., Cerutti, B., \& Nalewajko, K. 2018,
  MNRAS, 473, 4840, \dodoi{10.1093/mnras/stx2530}

\bibitem[{White {et~al.}(2020{\natexlab{a}})White, Dexter, Blaes, \&
  Quataert}]{White2020b}
White, C.~J., Dexter, J., Blaes, O., \& Quataert, E. 2020{\natexlab{a}}, ApJ,
  894, 14, \dodoi{10.3847/1538-4357/ab8463}

\bibitem[{White {et~al.}(2019)White, Quataert, \& Blaes}]{White2019}
White, C.~J., Quataert, E., \& Blaes, O. 2019, ApJ, 878, 51,
  \dodoi{10.3847/1538-4357/ab089e}

\bibitem[{White {et~al.}(2020{\natexlab{b}})White, Quataert, \&
  Gammie}]{White2020a}
White, C.~J., Quataert, E., \& Gammie, C.~F. 2020{\natexlab{b}}, ApJ, 891, 63,
  \dodoi{10.3847/1538-4357/ab718e}

\bibitem[{White {et~al.}(2016)White, Stone, \& Gammie}]{White2016}
White, C.~J., Stone, J.~M., \& Gammie, C.~F. 2016, ApJS, 225, 22,
  \dodoi{10.3847/0067-0049/225/2/22}

\bibitem[{Witzel {et~al.}(2012)Witzel, Eckart, Bremer, Zamaninasab,
  Shahzamanian, Valencia-S., Sch{\"o}del, Karas, Lenzen, Marchili, Sabha,
  Garcia-Marin, Buchholz, Kunneriath, \& Straubmeier}]{Witzel2012}
Witzel, G., Eckart, A., Bremer, M., {et~al.} 2012, ApJS, 203, 18,
  \dodoi{10.1088/0067-0049/203/2/18}

\bibitem[{Witzel {et~al.}(2018)Witzel, Martinez, Hora, Willner, Morris, Gammie,
  Becklin, Ashby, Baganoff, Carey, Do, Fazio, Ghez, Glaccum, Haggard,
  Herrero-Illana, Ingalls, Narayan, \& Smith}]{Witzel2018}
Witzel, G., Martinez, G., Hora, J., {et~al.} 2018, ApJ, 863, 15,
  \dodoi{10.3847/1538-4357/aace62}

\bibitem[{Yuan {et~al.}(2015)Yuan, Gan, Narayan, Sadowski, Bu, \&
  Bai}]{Yuan2015}
Yuan, F., Gan, Z., Narayan, R., {et~al.} 2015, ApJ, 804, 101,
  \dodoi{10.1088/0004-637X/804/2/101}

\bibitem[{Yuan \& Narayan(2014)}]{Yuan2014}
Yuan, F., \& Narayan, R. 2014, ARA\&A, 52, 529,
  \dodoi{10.1146/annurev-astro-082812-141003}

\bibitem[{Yusef-Zadeh {et~al.}(2015)Yusef-Zadeh, Bushouse, Sch{\"o}del, Wardle,
  Cotton, Roberts, Nogueras-Lara, \& Gallego-Cano}]{YusefZadeh2015}
Yusef-Zadeh, F., Bushouse, H., Sch{\"o}del, R., {et~al.} 2015, ApJ, 809, 10,
  \dodoi{10.1088/0004-637X/809/1/10}

\end{thebibliography}

\end{document}